\documentclass[useAMS,usenatbib]{mn2e}
\usepackage{graphicx,rotate,url,mathptmx,lscape,times}
\def\gtsim{\mathrel{\hbox{\rlap{\hbox{\lower4pt\hbox{$\sim$}}}\hbox{$>$}}}}
\def\lesssim{\mathrel{\hbox{\rlap{\hbox{\lower4pt\hbox{$\sim$}}}\hbox{$<$}}}}
\def\Msunpyr{M$_{\odot}\,$yr$^{-1}$}
\def\Msun{M$_{\odot}$}

\def\erg{{\rm\thinspace erg}}

\def\km{{\rm\thinspace km}}

\def\Msun{\hbox{$\rm\thinspace M_{\odot}$}}

\def\s{{\rm\thinspace s}}
\def\ps{{\rm\thinspace s^{-1}}}

\def\yr{{\rm\thinspace yr}}
\def\ergps{\hbox{$\erg\s^{-1}\,$}}
\def\ergpsphz{\hbox{$\erg\s^{-1}\,{\rm Hz}^{-1}$}}

\def\kmps{\hbox{$\km\ps\,$}}

\def\Msunpyr{\hbox{$\Msun\yr^{-1}\,$}}

\def\ha{\hbox{{\rm H}$\alpha$}}
\def\lya{\hbox{{\rm Ly}$\alpha$}}
\def\hb{\hbox{{\rm H}$\beta$}}

\def\oiii{\hbox{{\rm [O{\sc iii}]}}}

\def\nii{\hbox{{\rm [N{\sc ii}]}}}

\def\h0{\hbox{{\rm H}$^0$}}
\def\gband{$g_{475}$}
\def\iband{$I_{814}$}
\def\jband{$J_{110}$}
\def\hband{$H_{160}$}

\DeclareMathAlphabet{\vib}{OML}{cmm}{m}{it}
\begin{document}

\title[The growth and assembly of a massive galaxy]{The growth and assembly of a massive galaxy at $z\sim2$}

\author[N.~A.~Hatch et al.]
      {N. A. Hatch$^{1}$\thanks{E-mail:hatch@strw.leidenuniv.nl}, R.~A.~Overzier$^{2}$, 
      J.~D.~Kurk$^{3}$, G.~K.~Miley$^{1}$, H.~J.~A.~R\"ottgering$^{1}$  \newauthor  and  A.~W.~Zirm$^{4}$
      \\$^{1}$Leiden Observatory, University of Leiden, P.B. 9513, Leiden 2300 RA, The Netherlands\\$^{2}$Max-Planck-Institut f\"ur Astrophysik, Karl-Schwarzschild-Str.\,1, D-85748, Garching, Germany \\
$^{3}$Max-Planck-Institute f\"ur Astronomie, K\"onigstuhl 17, 69117 Heidelberg, Germany\\$^{4}$Department of Physics andAstronomy, Johns Hopkins University, 3400 North Charles Street, Baltimore, MD21218 
} 

\date{Accepted
      Received }

\pubyear{}

\maketitle

\label{firstpage}
\begin{abstract}
\noindent 
We study the stellar mass assembly of the Spiderweb Galaxy (MRC\,1138-262), a massive $z=2.2$ radio galaxy in a protocluster and the probable progenitor of a brightest cluster galaxy. Nearby protocluster galaxies are identified and their properties are determined by fitting stellar population models to their rest-frame ultraviolet to optical spectral energy distributions.  We find that within 150\,kpc of the radio galaxy the stellar mass is centrally concentrated in the radio galaxy, yet most of the dust-uncorrected, instantaneous star formation occurs in the surrounding low-mass satellite galaxies. We predict that most of the galaxies within 150\,kpc of the radio galaxy will merge with the central radio galaxy by $z=0$, increasing its stellar mass by up to a factor of $\simeq2$. However, it will take several hundred Myr for the first mergers to occur, by which time the large star formation rates are likely to have exhausted the gas reservoirs in the satellite galaxies. The tidal radii of the satellite galaxies are small, suggesting that stars and gas are being stripped and deposited at distances of tens of kpc from the central radio galaxy. These stripped stars may become intracluster stars or form an extended stellar halo around the radio galaxy, such as those observed around cD galaxies in cluster cores.

\end{abstract}

\section{Introduction}
 
High redshift radio galaxies (HzRGs) are among the most luminous and massive known galaxies (\citealt{Seymour2007} and references therein). They contain vast reservoirs of atomic and molecular gas, and are commonly embedded  in giant ionized gas halos. A number of well studied HzRGs have been found to be undergoing violent activity such as extreme star formation and merging. They contain active galactic nuclei (AGN) implying the co-evolution of their supermassive black holes and stellar hosts. Some radio galaxies have been found to be located in dense regions of the early Universe that may collapse to form clusters of galaxies with masses greater than $10^{14}$\Msun\ when extrapolated to $z=0$ \citep{Venemans2007}. Therefore HzRGs are likely to be the progenitors of low-redshift brightest cluster galaxies (BCGs; \citealt{MileydeBreuck2008,Zirm2005,Pentericci1997}). 

With a mass of $\sim10^{12}$\Msun, the radio galaxy MRC 1138-262 at z = 2.2 (Fig. 1; \citealt{Pentericci1997,Miley2006,Overzier2007}), also known as the \lq Spiderweb Galaxy',  is one of the most massive galaxies at $z\sim2$ \citep{Seymour2007}. It lies in a protocluster identified by an overdensity of \lya\ and \ha\ emitters, extremely red objects and sub-mm bright galaxies \citep{Pentericci2000,Kurk2004,Kodama2007,Stevens2003,Zirm2008}. The radio galaxy is 2\,mag brighter in $Ks$ than any other protocluster galaxy candidate within a 1.5\,Mpc physical radius \citep{Kurk2004}. It is therefore highly likely that this radio galaxy is the central galaxy within a massive dark matter halo at the centre of the protocluster, and that the nearby galaxies may become satellite galaxies, or eventually merge with it to form a brightest cluster galaxy. 

{\it Hubble Space Telescope} ({\it HST}) images of this radio source in \citet{Miley2006} show that MRC\,1138-262 is surrounded by tens of UV-bright clumps and galaxies. These observations agree qualitatively with predictions of hierarchical galaxy formation theory. In this work we examine the properties of the nearby galaxies in more detail, in order to quantify the future growth and mass assembly of the radio galaxy through the accretion of these galaxies. We combine high resolution rest-frame UV and optical data to identify protocluster galaxies that are likely to merge with the radio galaxy. By comparing their photometry to stellar population models, we estimate the stellar mass of these galaxies. We then use the observed stellar mass distribution as initial conditions to examine the future growth and assembly of the radio galaxy.

 In section \ref{observations} we describe the observations and data reduction and section \ref{selection} describes how the protocluster galaxies are identified.  Section \ref{fitting} gives details of the spectral energy distribution (SED) fitting, through which we derive the masses of the galaxies. These properties are discussed and compared to semi-analytic models in section \ref{results_section}.  In section \ref{discussion} we use simple analytic approximations of the merging timescale to examine the evolution of MRC\,1138-262. Throughout this work we use $H_0=71$\kmps, $\Omega_M=0.27$, and $\Omega_\Lambda=0.73$ \citep{Spergel2003}, and at the redshift of MRC\,1138-262  (z=2.156) the linear scale is 8.4\,kpc\,arcsec$^{-1}$. All magnitudes are AB magnitudes. 

\section{Observations}
\label{observations}
The protocluster surrounding MRC\,1138-262 was observed with the {\it Advanced camera for Surveys} (ACS) instrument on the {\it HST} though the F475W (\gband) filter for 9 orbits, and 10 orbits  though the F814W (\iband) filter. The data were reduced using the ACS pipeline ({\sc apsis}; \citealt{Blakeslee2003}). Details of the data reduction are given in \citet{Miley2006} and \citet{Hatch2008}. The protocluster was also observed with NICMOS camera 3 on {\it HST} for 3 orbits though each of the F110W (\jband) and F160W (\hband) filters. Further discussion of the NICMOS observations can be found in \citet{Zirm2008}. 

The {\it HST} observations were supplemented with a 3\,hour exposure in $U_{n}-$band from the LRIS-B instrument on the Keck telescope (P.I. W. van Breugel), and near-infrared (near-IR) images from the Very Large telescope (VLT) ISAAC instrument: $Ks-$band (1.6\,hour) and a narrow-band image centred on 2.07$\micron$ (4.8\,hour). The $U_{n}-$band data has a  full width at half maximum (FWHM) of 1.2\,arcsec, and both near-IR images have a FWHM of 0.45\,arcsec, measured from bright, unsaturated stars that lie close to the radio galaxy. 
Details of the reduction of the near-IR observations can be found in \citet{Kurk2004}. \ha\ emission from the protocluster at z$=2.156$ lies within the bandwidth of the narrow-band NB207 filter.  Both an \ha\ emission-line image of the protocluster and an emission-line-free $Ks$ continuum image were created from the near-IR dataset. 

The images were registered and matched in pixel-scale  (0.2\,arcsec\,pixel$^{-1}$) and resolution to the $Ks$ image. The 3$\sigma$ image depths of the images are 27.7\,mag in \gband, 27.7\,mag in  \iband, 26.2\,mag in \jband, 26.6\,mag in \hband, 24.4\,mag in $Ks$ and 26.8\,mag in $U_n$.
Magnitudes were measured using SExtractor \citep{Bertin1996} in double-image mode, using a "white-light" image as a detection image. The detection image was created by combining the high signal-to-noise \gband, \iband, \jband\ and \hband\ {\it HST} images.  Regions with 7 or more connecting pixels which had a signal-to-noise ratio greater than 4 times the local background noise  were identified and isophotal magnitude were measured. 

The point spread function (PSF) of the $U_n-$band data is more than twice as large as  the PSF of the other images. To preserve the high resolution of the {\it HST} and $Ks$ images we scale the $U_n$ magnitudes by a factor to account for the light dispersed beyond the galaxy isophotal detection area by the larger PSF. The \gband\ was smoothed with a Gaussian of FWHM 1.2\,arcsec to match the resolution of the $U_n$ image. Magnitudes were obtained from both the \gband\ and $U_n$ images using the original detection image and multiplied by the ratio of flux within the isophotal area of the 0.45\,arcsec resolution \gband\  image to the 1.2\,arcsec resolution \gband\  image. Objects which are fainter in the 0.45\,arcsec resolution image than the 1.2\,arcsec resolution \gband\  image are blended and no $U_n$ magnitude was assigned to these objects. All magnitudes were corrected for Galactic extinction of $A_{B}=0.172$ \citep{Schlegel1998}.

\subsection{Emission-line contribution}
Ly$\alpha$ emitted from galaxies in the protocluster lies within the bandwidth of the $U_n$ filter, although the filter transmission at this wavelength is only 30\% of the maximum filter transmission. \citet{Kurk2004} lists the bright \lya\ emitters close to the radio galaxy.  Galaxy 10 (see Fig.~\ref{overlay}) has an equivalent width (EW) greater than 1000\AA. The $U_n$ magnitude for this galaxy is likely to be dominated by the \lya\ emission. We therefore assume the observed magnitude is an upper limit on the $U_n$ continuum emission. The other \lya\ emitters have EWs of 30-100\AA, therefore a maximum of 5\% of the emission in the $U_n-$band results from \lya\ line emission and we do not include any correction for this. 

Both the C{\sc iv} and He {\sc ii} emission lines fall within the \gband\ bandwidth, and contribute greatly to the diffuse intergalactic light \citep{Hatch2008}, but spectroscopy shows that neither line is detected in the majority of the \lya\ emitting galaxies close to MRC\,1138-262 \citep{Kurkthesis}. Only the radio galaxy and galaxy 10 have large combined C{\sc iv} and He {\sc ii} EWs of 101\AA\ and 87\AA\ respectively and the \gband\ magnitudes are reduced by 22.5\% and 20\% respectively. 

H$\beta$ and \oiii\ emitted from the protocluster galaxies falls within the \hband\ bandwidth. The contribution in \hband\ from these lines is estimated using the \ha\ image and observations of MRC\,1138-262 from the integral field unit spectrometer SINFONI, which show that \ha+\nii$>2.4\times$\oiii+\hb\ (N.\,Nesvadba -- private communication). The line emission from the galaxies is likely to be fainter than the line emission from the halo, therefore a maximum of 5\% of the \hband\ flux can be due to line emission. As we do not know exactly how much line emission should be removed from each galaxy, we do not remove this component. The \ha\ emission was removed from the $Ks$ image using the narrow-band NB207 image.

\subsection{Removing the nuclear contribution}
The nuclear radio emission is associated with galaxy 1 and we refer to this galaxy as the radio galaxy (see Fig.\,2; \citealt{Pentericci1997}). To remove the point source contribution from the active galactic nucleus, all galaxies, except for the radio galaxy, are masked out from the original high-resolution {\it HST} images, and the resulting image is fitted with an ellipsoid model to obtain the best-fitting centroid, position angle and axial ratio. The azimuthally averaged radial profile is fitted with a classical de Vaucouleurs profile plus a point source smoothed to the resolution of the image. The centroid of the point source is not constrained to lie at the centre of the large-scale ellipsoid. The point source contribution to the   $Ks$, \hband, \jband\ and \iband\ images is 15\%, 30\%, 23\%, 9\%  respectively and the galaxy profile in \gband\ and $Un-$bands is consistent with no point source contribution. Most AGN have nuclear starbursts and therefore a small percentage of star formation is likely being removed together with the point source.

The radio galaxy radial profile in \jband\ and \hband\ is well described by a  central point source plus a de Vaucouleurs profile with a characteristic radius of 0.6\,arcsec (5\,kpc -- comparable to low redshift cD and giant Elliptical galaxies). However, in the rest-frame UV light (\iband\ and \gband)  the radio galaxy is not well described by the de Vaucouleurs plus point source profile at large radii because of the large contribution of extended diffuse light (see \citealt{Hatch2008}). 
\begin{figure}
\centering
\includegraphics[width=0.95\columnwidth]{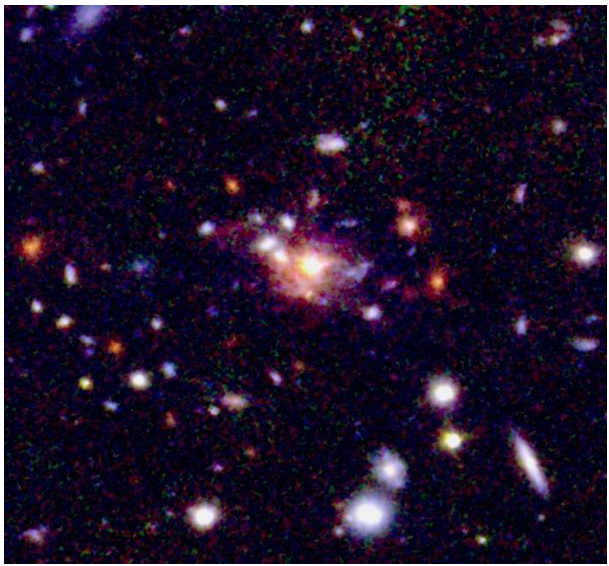}\caption{Combined \iband-\jband-$K_{s}$ image of MRC\, 1138-262 (centre) and its neighbouring satellites galaxies within 150\,kpc. Total extent of the image is 300\,kpc on each side. Galaxies identified to be within the protocluster are marked and labelled in Fig.\,\ref{overlay}. Note the bridge of red light connecting galaxies 17 and 18 possibly arising from tidal stripping.\label{full_colour}}
\end{figure}

\begin{figure}
\centering
\includegraphics[width=1.0\columnwidth, angle=90]{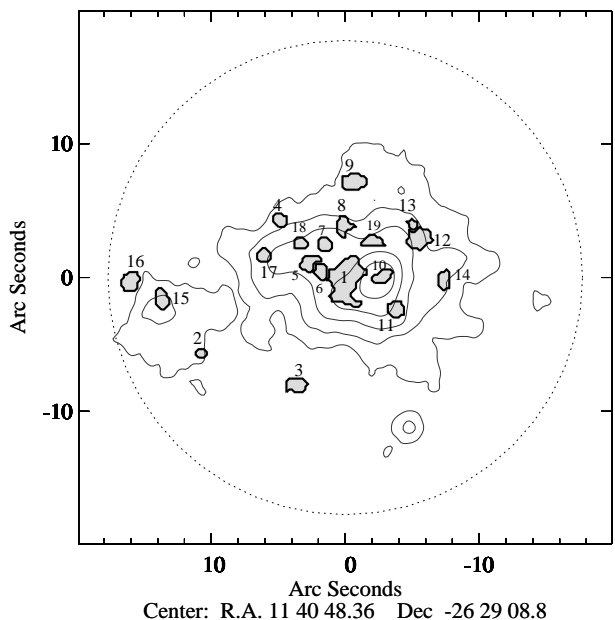}\caption{Galaxies selected to be at the same redshift as the radio galaxy are outlined in black contours. Numbers correspond to galaxy IDs in Table \ref{results} and Fig.\,\ref{SED_fits}. The \lya\ halo of the radio galaxy is shown in grey contours. The red galaxies with little current star formation (galaxies 2, 4 and 16)  are located towards the edge of the \lya\ halo, whilst those with large star formation rates are located close to the radio galaxy. The dotted circle encompasses the area within 150\,kpc of the centre of the radio nucleus.
\label{overlay}}
\end{figure}

\section{Selecting protocluster galaxies}
\label{selection}
We wish to study the central region of the protocluster, the radio galaxy and the surrounding galaxies that will interact or merge with the radio galaxy. To minimize the contamination of more distant protocluster galaxies, only galaxies within a radius of $\sim$150\,kpc (18\,arcsec) of the radio galaxy were selected. This is the observed maximum radial extension of the \lya\ halo which surrounds the radio galaxy and corresponds to the virial radius of a $\sim10^{13}$\Msun\ object (see Fig.~\ref{overlay}). We detect 64 separate objects within this area.

%\subsection{Emission-line galaxies}

Eight of the 64 objects are confirmed to be protocluster galaxies through the detection of \lya\ emission (galaxies 1, 5, 6, 7, 9, 10, 11, 15; \citealt{Kurkthesis,Kurk2004}; see Fig.~\ref{overlay}). Five objects are observed to emit \ha\ emission, galaxies 1,10,11,12 are identified by \citet{Kurk2004} and we identify a further galaxy (ID 8) at 11:40:48.37 $-$26:29:05.2 (J2000) with an EW of 153\AA. In total 9 satellites of the radio galaxy are identified by line emission detections. 

%\subsection{Selection of red protocluster galaxies}
Selecting protocluster galaxies through line emission alone biases the sample toward star forming galaxies.  At the redshift of the protocluster the Balmer and 4000\AA\ break falls within the \jband\ filter, therefore we can make a simple colour cut to select candidate red protocluster galaxies. \citet{Zirm2008} show that there is a significant overdensity of galaxies in this protocluster field with colours $1.1<$\jband--\hband$<2.1$. Using the same colour cut we identify 7 red galaxies (galaxies 1, 2, 4, 12, 13, 14 and 16) two of which are also \ha\ emitters. 

We include 3 additional galaxies in our catalogue of satellite galaxies based on their proximity to the radio galaxy and their peculiar morphologies. Galaxies 17 and 18 have a bridge of red light connecting them which corresponds to a surface brightness enhancement in the \lya\ halo (see Fig.\ref{full_colour} and \ref{overlay}) suggesting that they are an interacting system, but the photometric redshift estimation code {\sc bpz}  \citep{BPZ} assigns redshifts of 3.6 and 0.3 respectively with low probability. Galaxy 19 has a peculiar linear morphology and lies close in projection to the radio galaxy. A concentration of chain and tadpole galaxies were found around the radio galaxy \citep{Miley2006} suggesting that they are located in the protocluster core. We include galaxies 17, 18 and 19 in the subsequent analysis but note that these galaxies have not been formally identified with the protocluster.

%\subsection{Photometric redshifts}
Photometric redshifts of all objects were obtained using the bayesian photometric redshift estimation code {\sc bpz} \citep{BPZ}, utilizing the CWWSB template set as described in \citet{BPZ} and 2 additional simple stellar population models with ages of 5  and 25\,Myr \citep{Coe2006} from \citet{BC03}. We used a prior based on the galaxy population of the Northern Hubble Deep Field and imposed an additional weak prior of a cluster at z=2.2 with 1\% of the galaxies in the field lying within the cluster. This is a weak additional prior since 10\% of the galaxies in the 0.283\,arcmin$^{2}$ field are identified as protocluster galaxies from line emission.

The \ha\ emitters labelled 8 and 12, and all protocluster galaxies selected by a red colour-cut have photometric redshifts within $\Delta$z=0.3 of the redshift of the radio galaxy.  One further galaxy (3) has a photometric redshift of 2.2 with Bayesian Odds $O>0.9$. Including the cluster prior shifted the best fit redshifts of galaxies 13, 14 and 16 from $z\sim1.4$ to $z=2.2$, although $z=2.2$ lay within the 1$\sigma$ confidence intervals whether the cluster prior was included or not. Increasing the cluster prior such that 50 per cent of the galaxies in the field lie within the cluster had no further effect and did not increase the number of galaxies with photometric redshifts within $\Delta z=0.3$ of 2.2. The photometric redshifts of the blue galaxies could not be constrained using BPZ. Galaxies with high equivalent width of \lya\ generally have smaller UV colours across the Lyman break than galaxies with no \lya\ emission. Also, although the selection
based on U-G is a powerful technique of isolating \lq BX' galaxies at $z\sim2.2\pm0.3$ (see \citealt{Reddy2005}), our U band data does not reach deep enough to constrain the colour across the Lyman break accurately. \\

\noindent Within 150\,kpc of the radio galaxy, we find a total of 19 galaxies in the protocluster. It should be noted that the number of galaxies seen in the \iband\ images of \citet{Miley2006}  and \citet{Hatch2008} is greater than presented in this work. The lower resolution of the data used in this work removes many of the smaller clumps of bright emission. The large number of the small clumps seen near the radio galaxy in the \iband\ and \gband\ images suggest they are located in the same region as the radio galaxy although no redshift can be determined. Thus the 19 detected galaxies should be considered the observable subset of a larger population of protocluster galaxies.

\section{Stellar population synthesis modeling}
\label{fitting}
A number of works have derived the stellar population properties of $z\sim2$ galaxies using stellar population synthesis techniques. We follow a procedure similar to that of \citet{Papovich2001,Shapley2005} and \citet{Kriek2008} and determine stellar population properties by fitting the broad-band photometry with stellar population models. Our aim was to determine the age, intrinsic star formation rate, $E\rm{(}B-V\rm{)}$, star formation history and stellar mass of each galaxy within 150\,kpc of the radio galaxy.

We compare the spectral energy distribution (SED) of the galaxies with stellar population templates created using \citet{BC03} models. We used models with solar metallicity, and a \citet{Chabrier2003} initial mass function (IMF) extending from 0.1 to 100\Msun. We use the Padova 1994 stellar population evolution tracks to form single exponentially declining star formation histories with characteristic e-folding times ($\tau$) of 1, 0.5, 0.1, 0.05, and 0.01\,Gyr, and a model with a constant star-forming history. We consider models with 31 ages between $10^6$\,yrs since the onset of star formation and 3.0\,Gyr, the age of the Universe at $z=2.156$, the redshift of MRC\,1138-262; and 21 extinction values between $A_{V}=0$ and 4\,mag, assuming the Calzetti reddening law.  We compare the photometry to the stellar population synthesis templates and  determine the best-fit parameters by minimum-$\chi^2$ techniques.
Confidence intervals due to the photometric errors are derived by constructing 500 synthetic realisations of each galaxy, varying the photometry by a random amount drawn from the normal photometric error distribution. The stellar population models are compared to the synthetic data sets to derive new best-fitting models.  1$\sigma$ confidence intervals of the best fit parameters are derived from the minimum and maximum values enclosing 68 per cent of the simulations. We also derive best-fit parameters without the $U_n$--band information by comparing the \gband\ through $Ks-$band photometry to the stellar population models. The results of all the alternative fits are within the errors given in Table \ref{results}.

Stellar templates created with single exponentially decreasing or constant star formation histories provide lower limits to the galaxy mass, particularly if the galaxy has a very young age and active star formation. Such galaxies have relatively small mass-to-light ratios as the observed light is dominated by bright young stars, which may hide an older stellar population, that could potentially be more massive than the brighter young population. To estimate the amount of mass that may exist in an older stellar population we fit the photometry with a two component star formation history. One component is fixed to be a 3\,Gyr population with $\tau=0.01$\,Gyr, whilst the other model is fixed to be a 1\,Myr old constant star forming population. The galaxy mass estimated from this two-component model is an upper limit on the true stellar mass. 1$\sigma$ confidence intervals are drawn from 500 Monte Carlo simulations as described above.

The best-fit parameters depend systematically on the choice of metallicity and IMF, for example a Salpeter IMF gives mass-to-light ratios that are systematically higher by $\sim1.5$ than the Chabrier IMF.  Larger metallicities result in redder SEDs, and thus allowing the metallicity to decrease, can increase the best-fit extinction, and decrease the age of the stellar population. A full discussion of how the best-fit galaxy parameters depend on the IMF and the metallicity is given in \citet{Papovich2001}.  

The parameters derived from SED fitting are subject to large systematic uncertainties and degeneracies and we refer to \citet{Shapley2005} for details. The best-fit age, intrinsic star formation rate and extinction strongly depend on the assumed star formation history. The data are unable to constrain the star formation histories across our range of parameter space, for all the selected galaxies.  We therefore infer that the derived age, extinction and intrinsic star formation rate are unreliable, and will not be discussed further. In the following discussion we refer only to the derived stellar mass (columns 3 and 4 of Table \ref{results}), and the dust-uncorrected star formation rate (column 2), determined using the observed \gband\ luminosity and
\begin{equation}
SFR(\Msunpyr)=\frac{L_{1500}}{1.5\times10^{28}}(\ergpsphz)
\label{sfrg}
\end{equation}
\citep{Salim2007}.

Stellar masses do not suffer the same large systematic uncertainties due to the unconstrained star formation history \citep{Shapley2005} and therefore the  derived masses are likely to be more robust. We note that \citet{Shapley2005} use near-infrared photometry, which means the derived masses are more reliable. The rest-frame $V-$band is longest wavelength in this work, and therefore we expect larger uncertainties. Although {\it Spitzer} mid-infrared images are available, the high density of the galaxies in this region, and the low resolution of the data, does not allow us to robustly measure rest-frame near-infrared magnitudes of the individual galaxies.

Recent work has focused on the contribution to the rest-frame near-infrared light of thermally pulsating asymptotic giant branch stars, which occurs 0.2--2\,Gyrs after the onset of star formation \citep{Maraston2006,BC03}. The reddest band we use in this work is equivalent to the rest-frame $V$, and the contribution of thermally pulsating asymptotic giant branch stars at this wavelength is not important. We have compared our results from the \citet{BC03} models to the new Charlot \& Bruzual (2007) models which include the thermally  pulsating asymptotic giant branch stars. We find no significant change in the results given in Table \ref{results}, in particular the stellar masses are within the 1$\sigma$ errors quoted. 

We note that masses derived from multiband photometry are subject to systematic uncertainties from the assumed IMF  \citep{Conroy2008}. The uncertainty on the assumed IMF may drastically change the derived masses, possibly by more than a factor of 10. However, the relative masses of galaxies, which are of particular importance to this work, are derived with much greater certainty.

\section{Results}
\label{results_section}

\subsection{Properties of the galaxies}
\label{properties}

\begin{figure*}

\includegraphics[width=0.5\columnwidth]{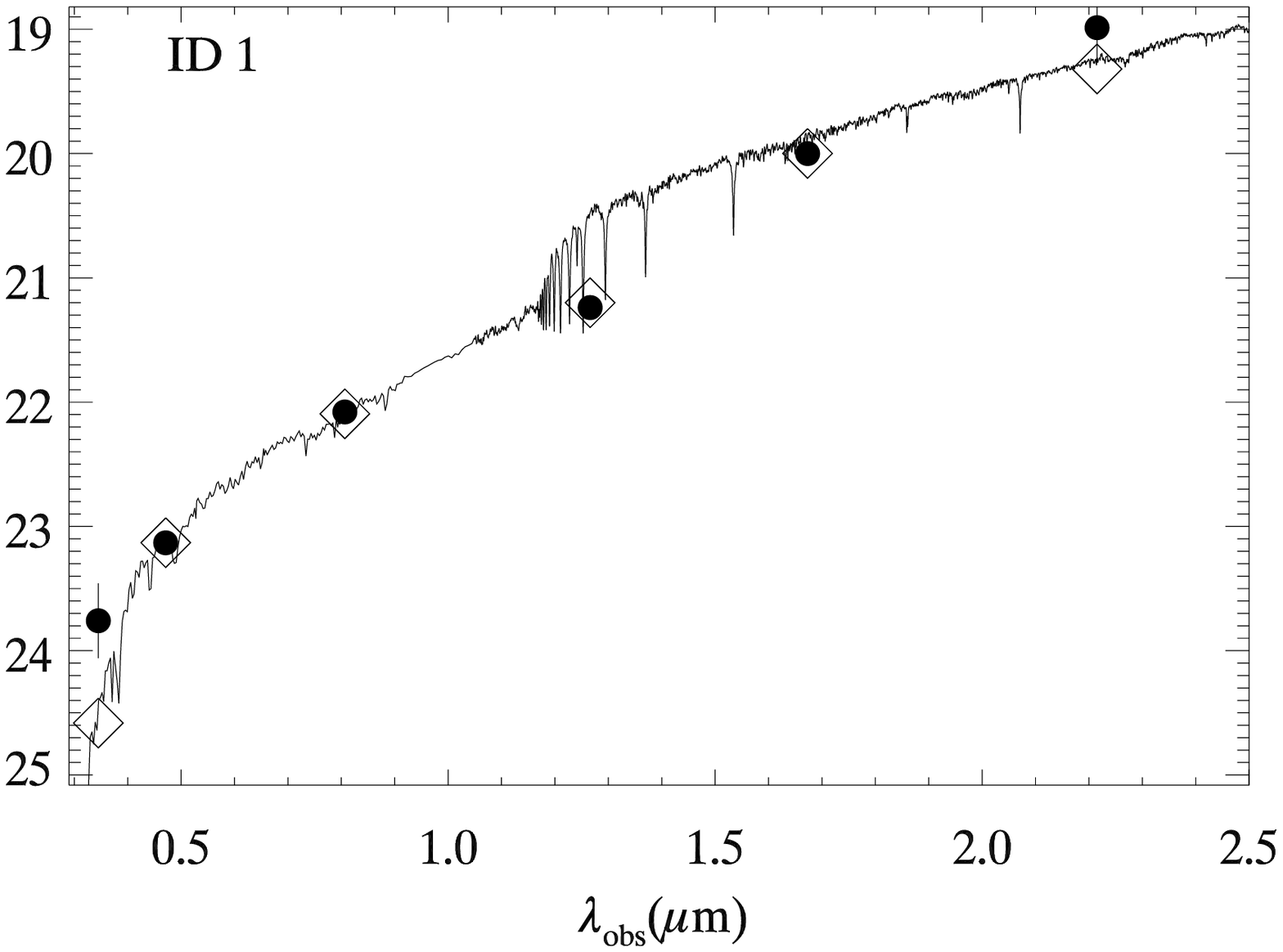}
\includegraphics[width=0.5\columnwidth]{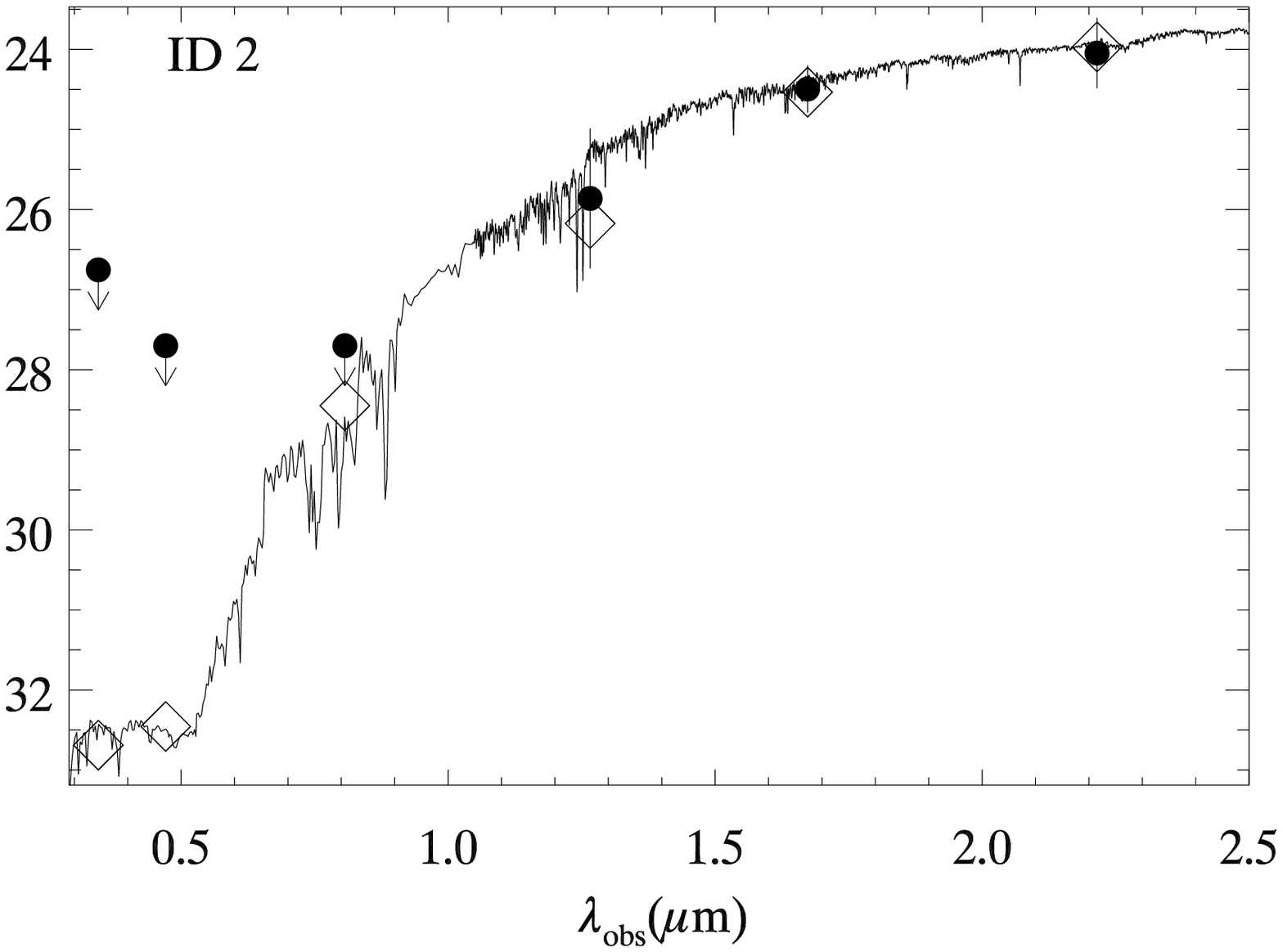}
\includegraphics[width=0.5\columnwidth]{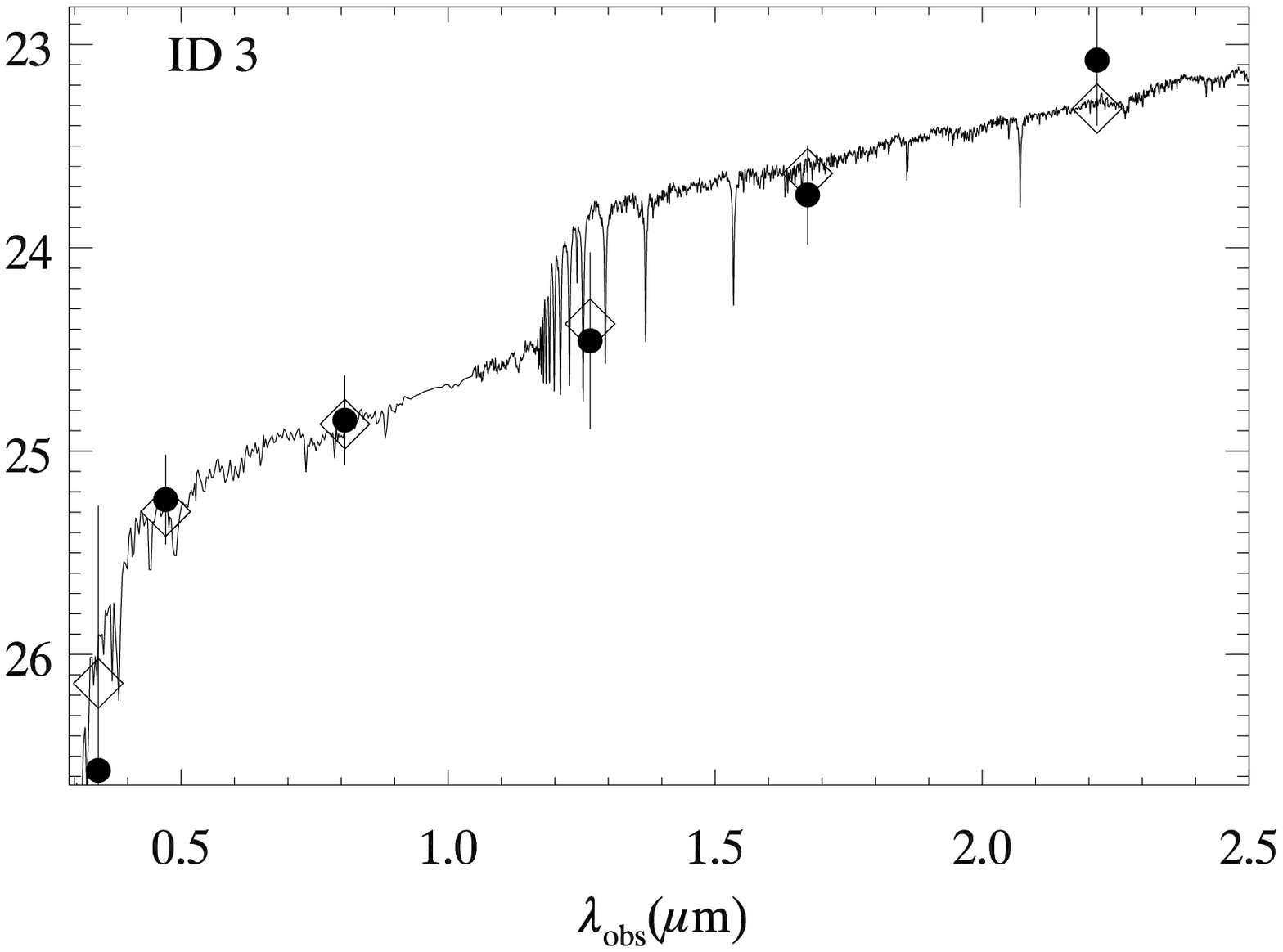}
\includegraphics[width=0.5\columnwidth]{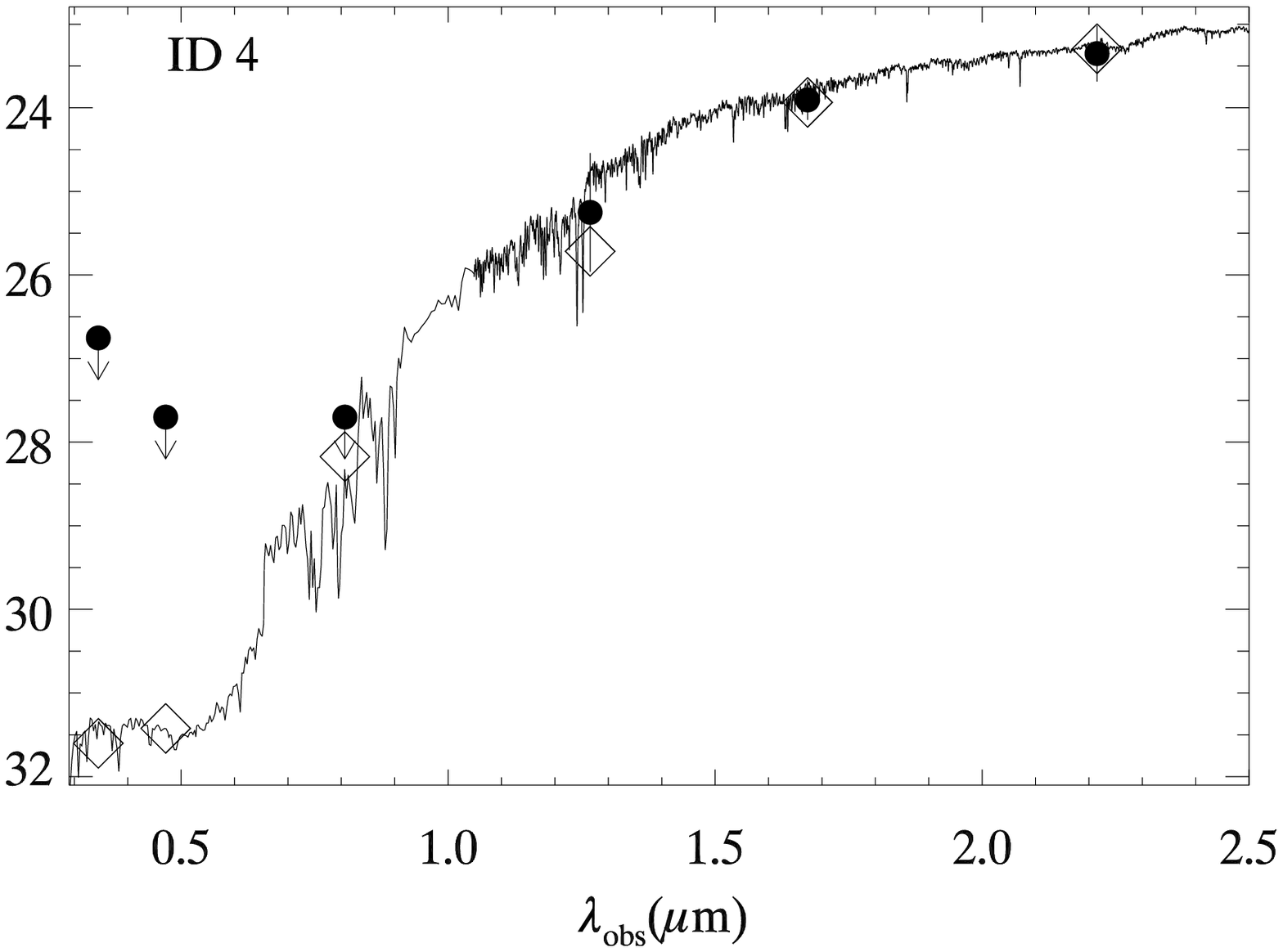}
\includegraphics[width=0.5\columnwidth]{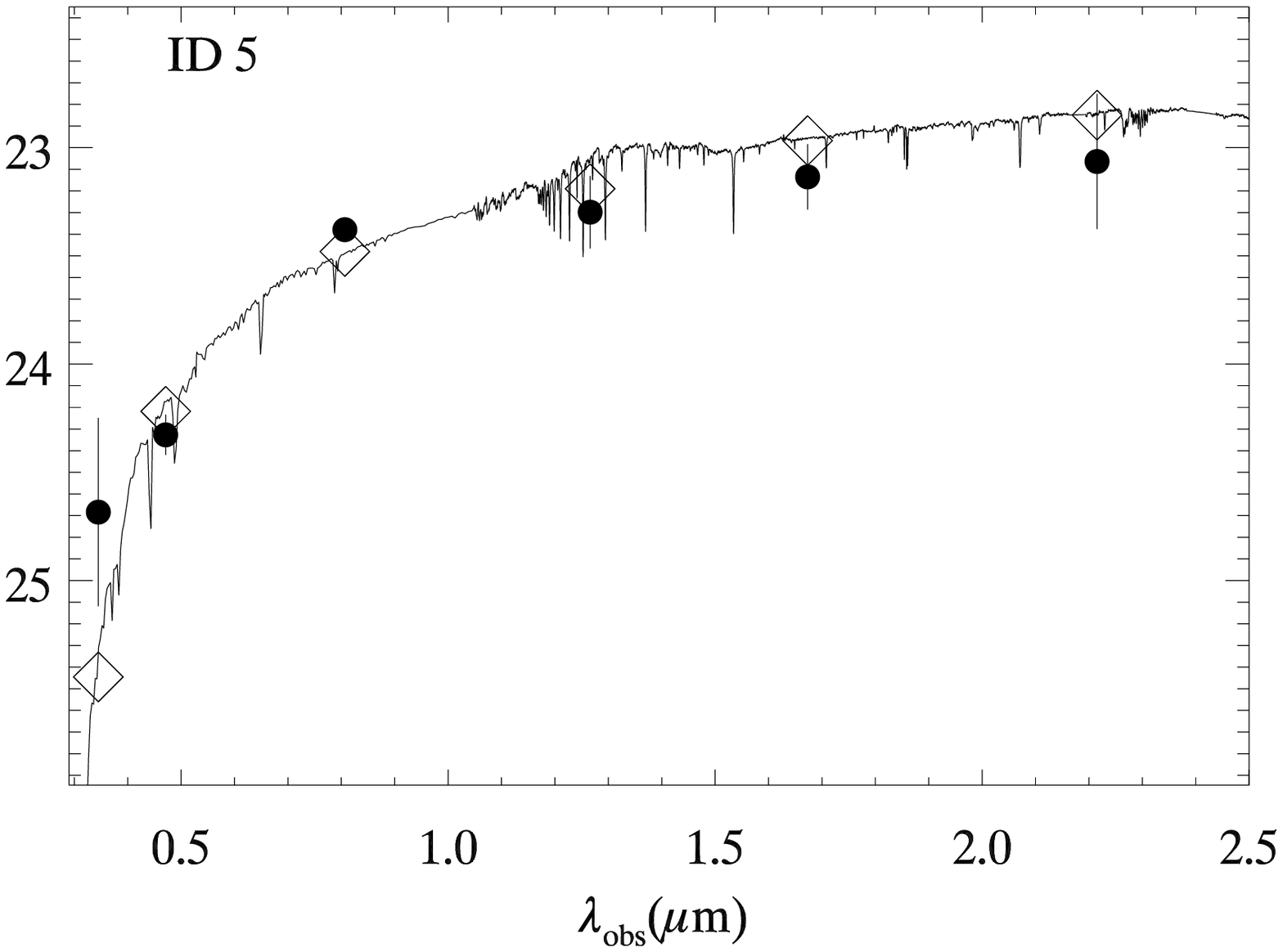}
\includegraphics[width=0.5\columnwidth]{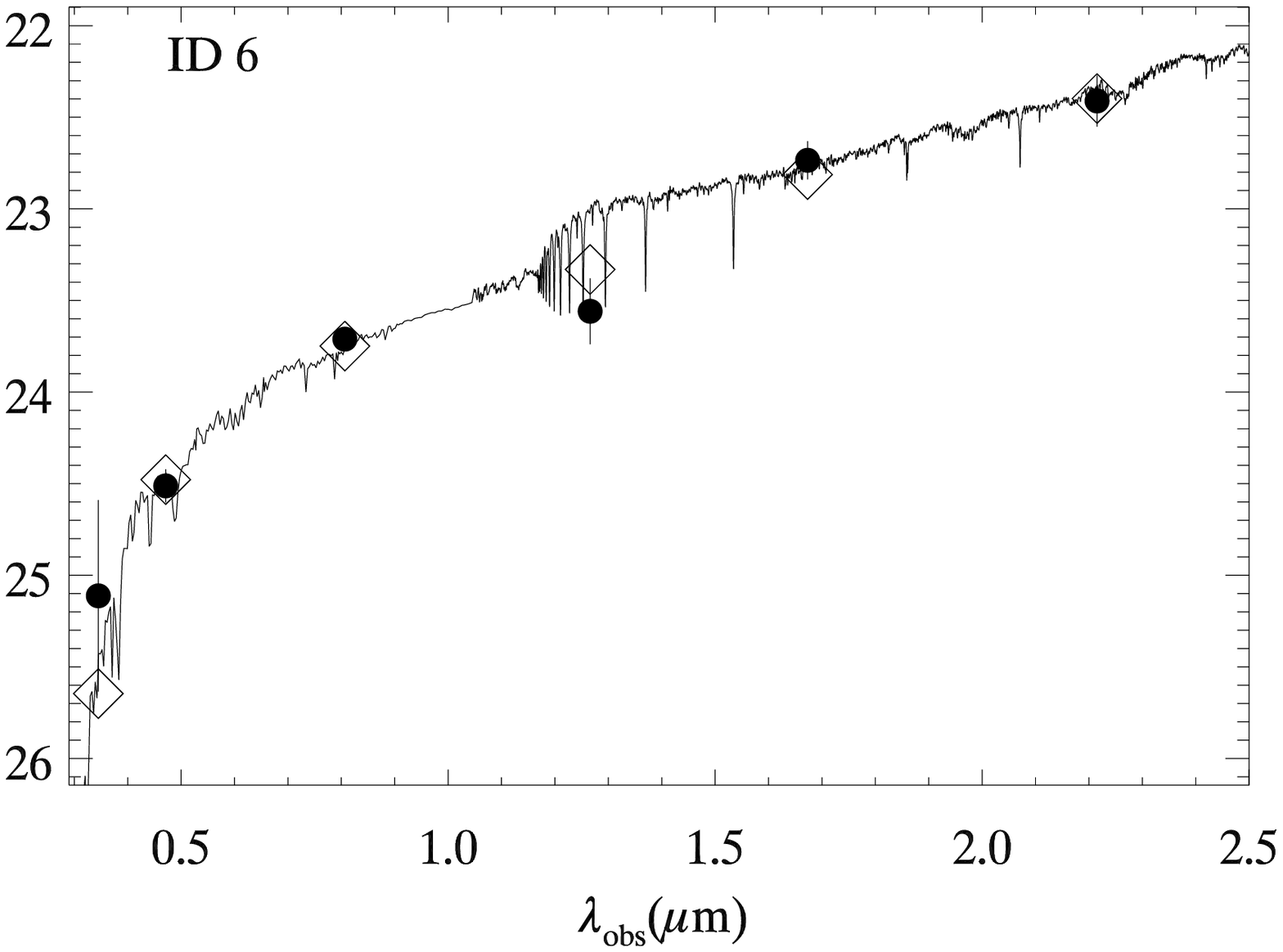}
\includegraphics[width=0.5\columnwidth]{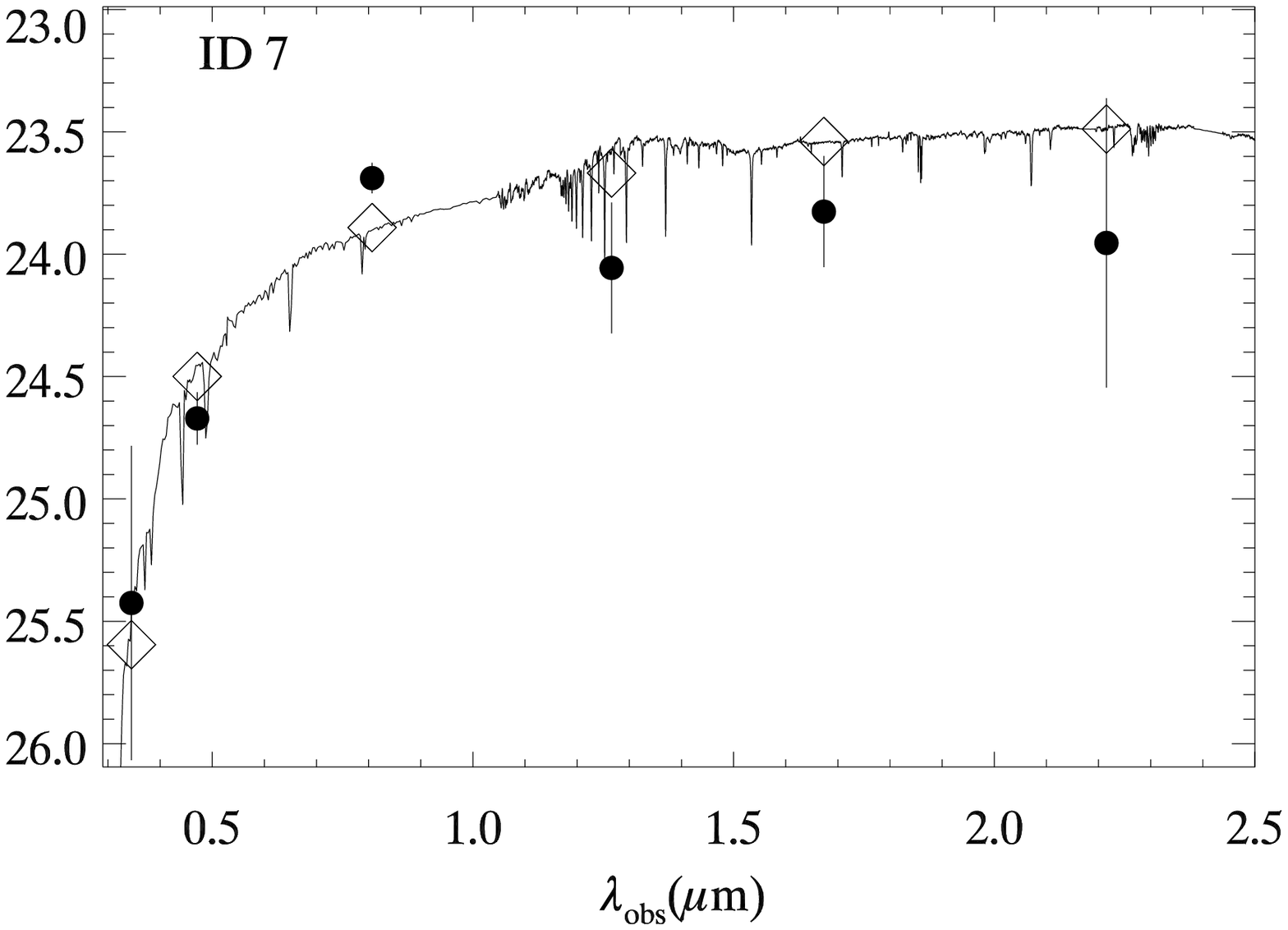}
\includegraphics[width=0.5\columnwidth]{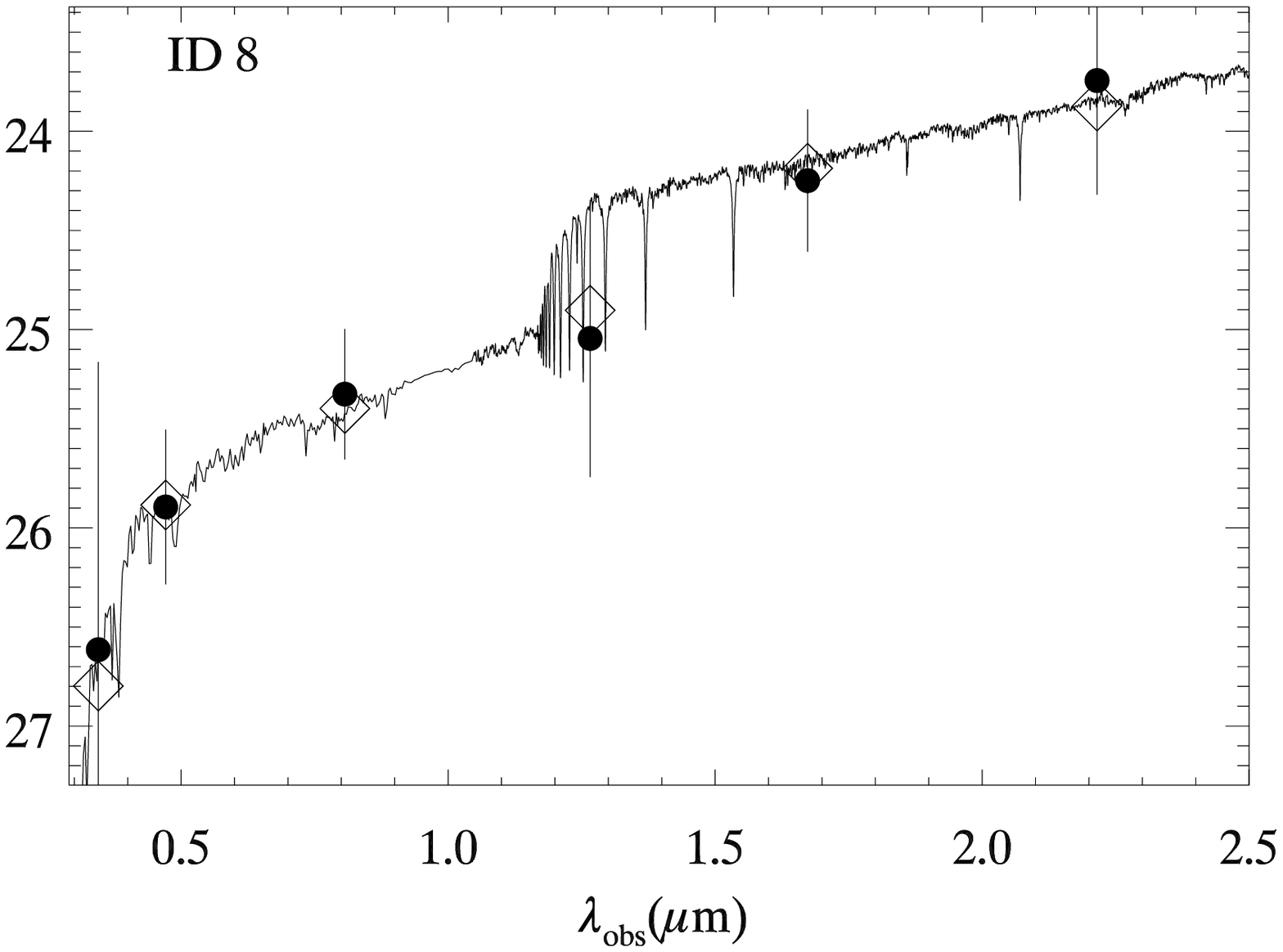}
\includegraphics[width=0.5\columnwidth]{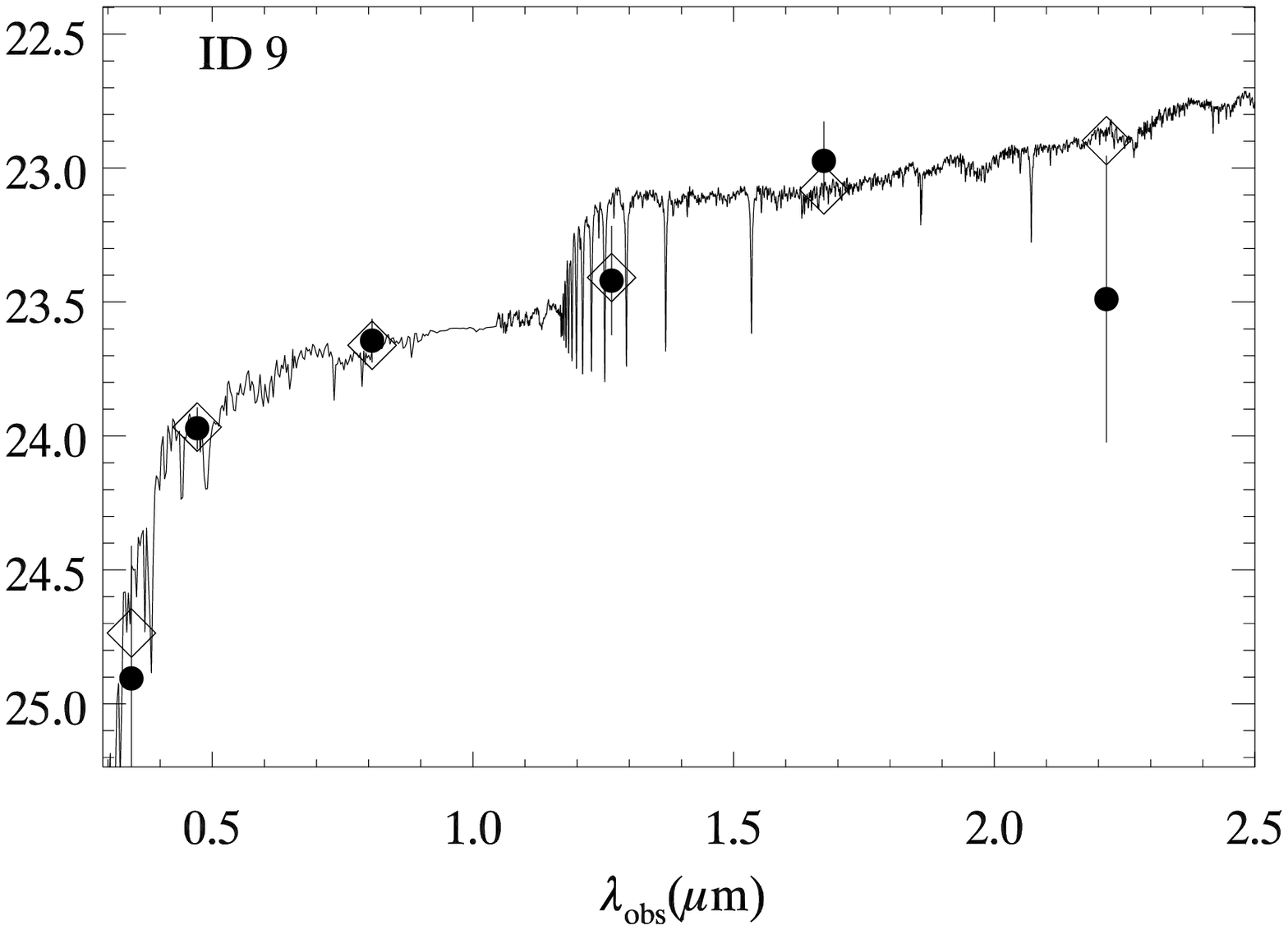}
\includegraphics[width=0.5\columnwidth]{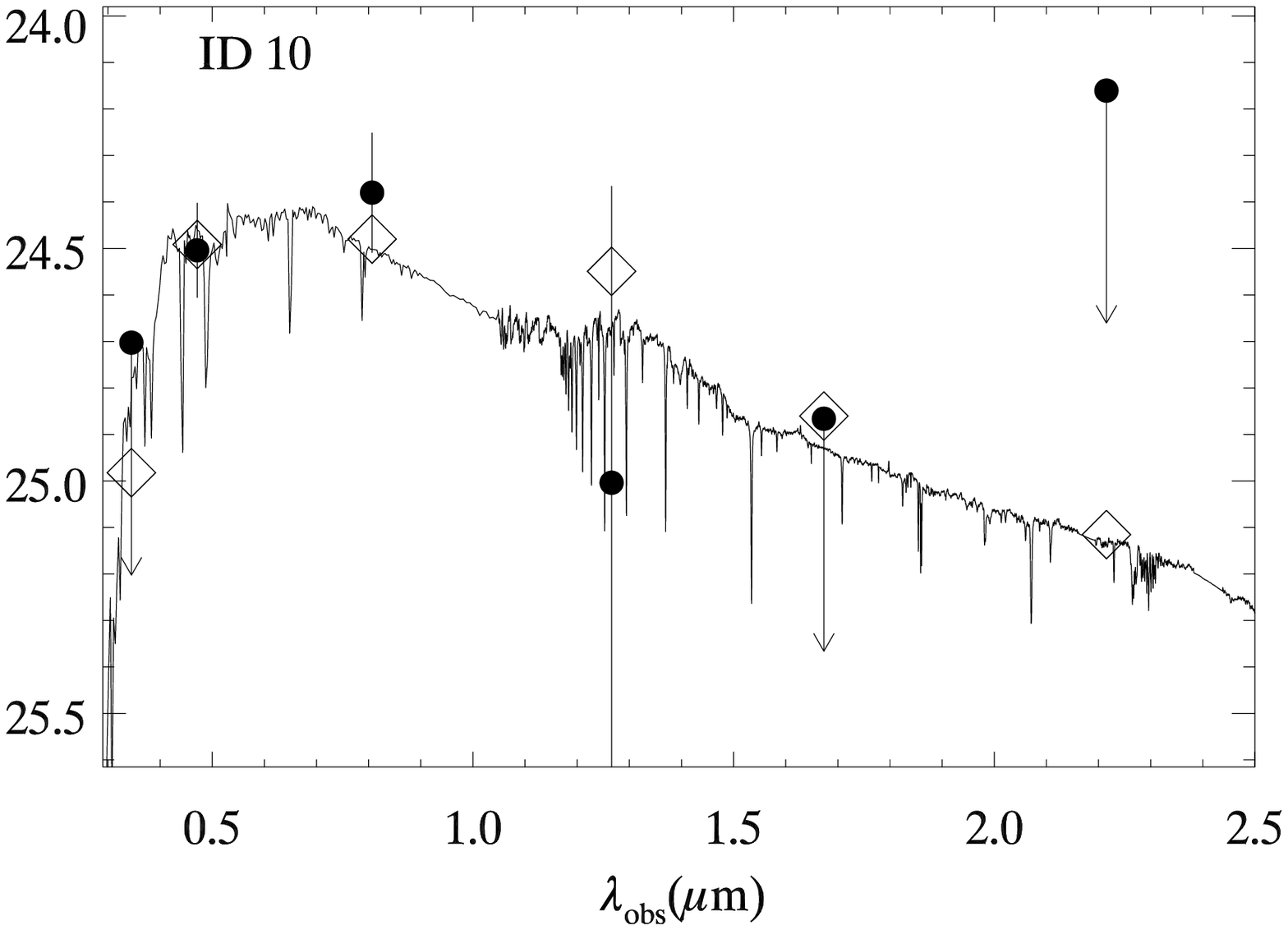}
\includegraphics[width=0.5\columnwidth]{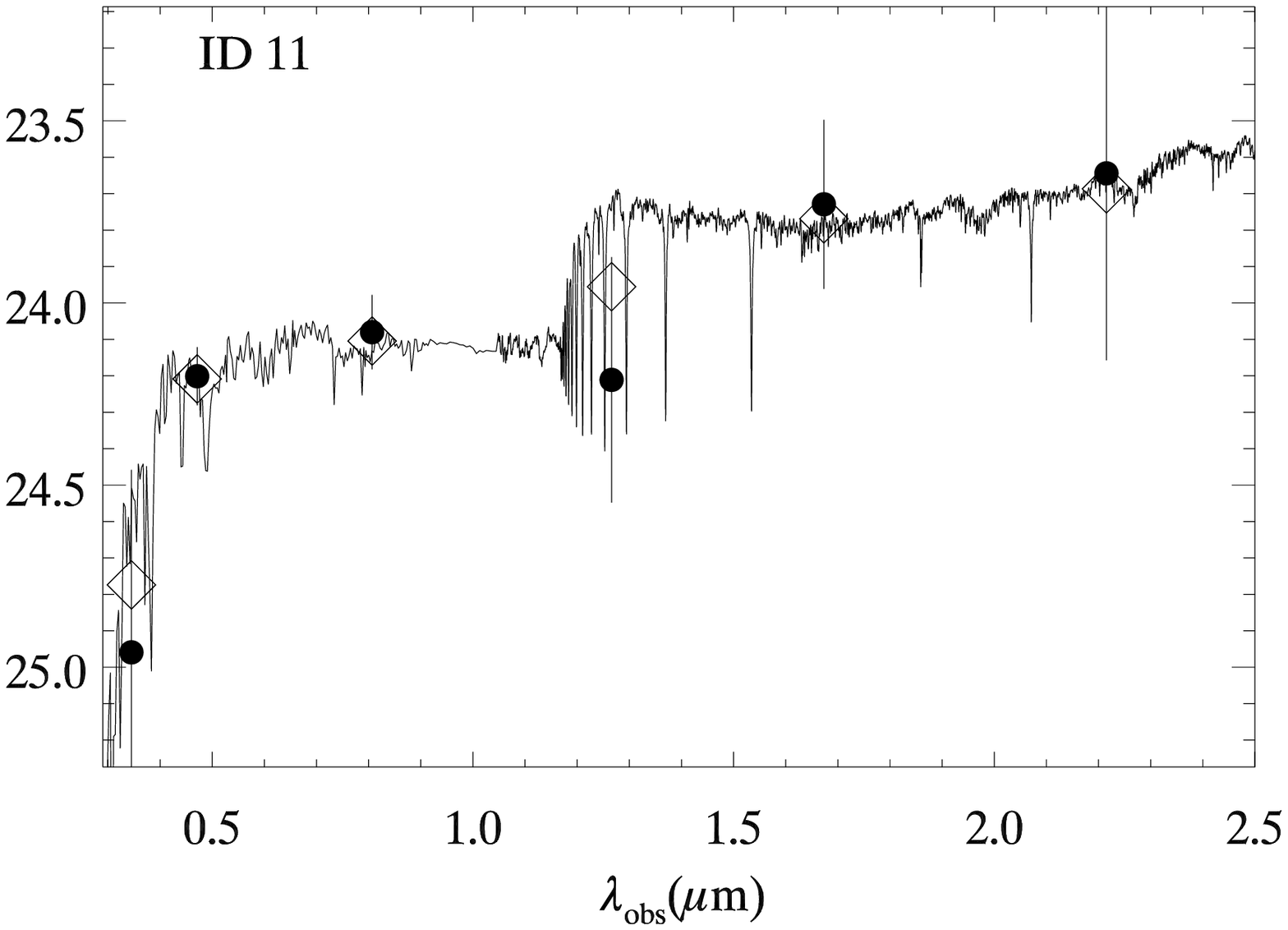}
\includegraphics[width=0.5\columnwidth]{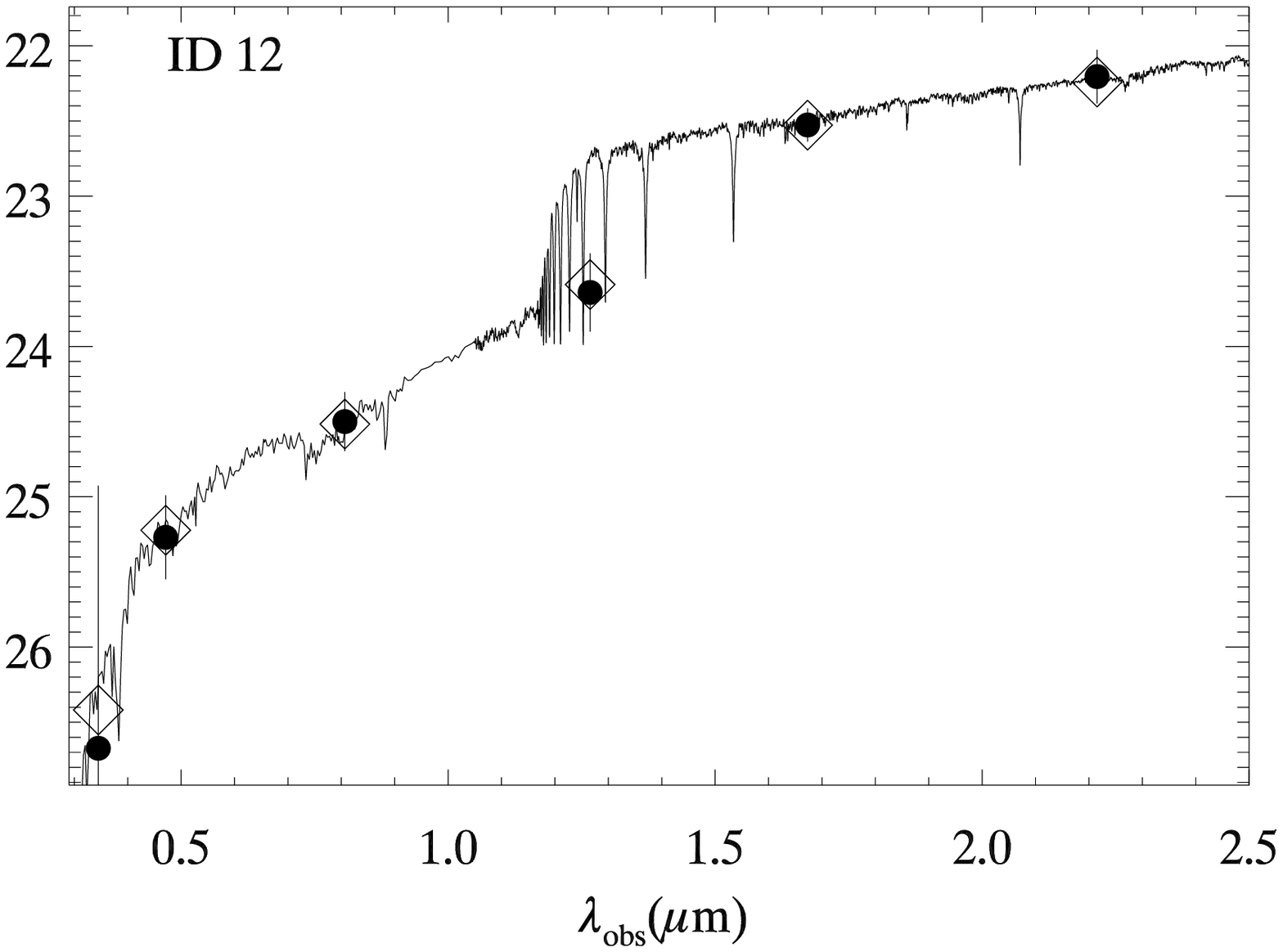}
\includegraphics[width=0.5\columnwidth]{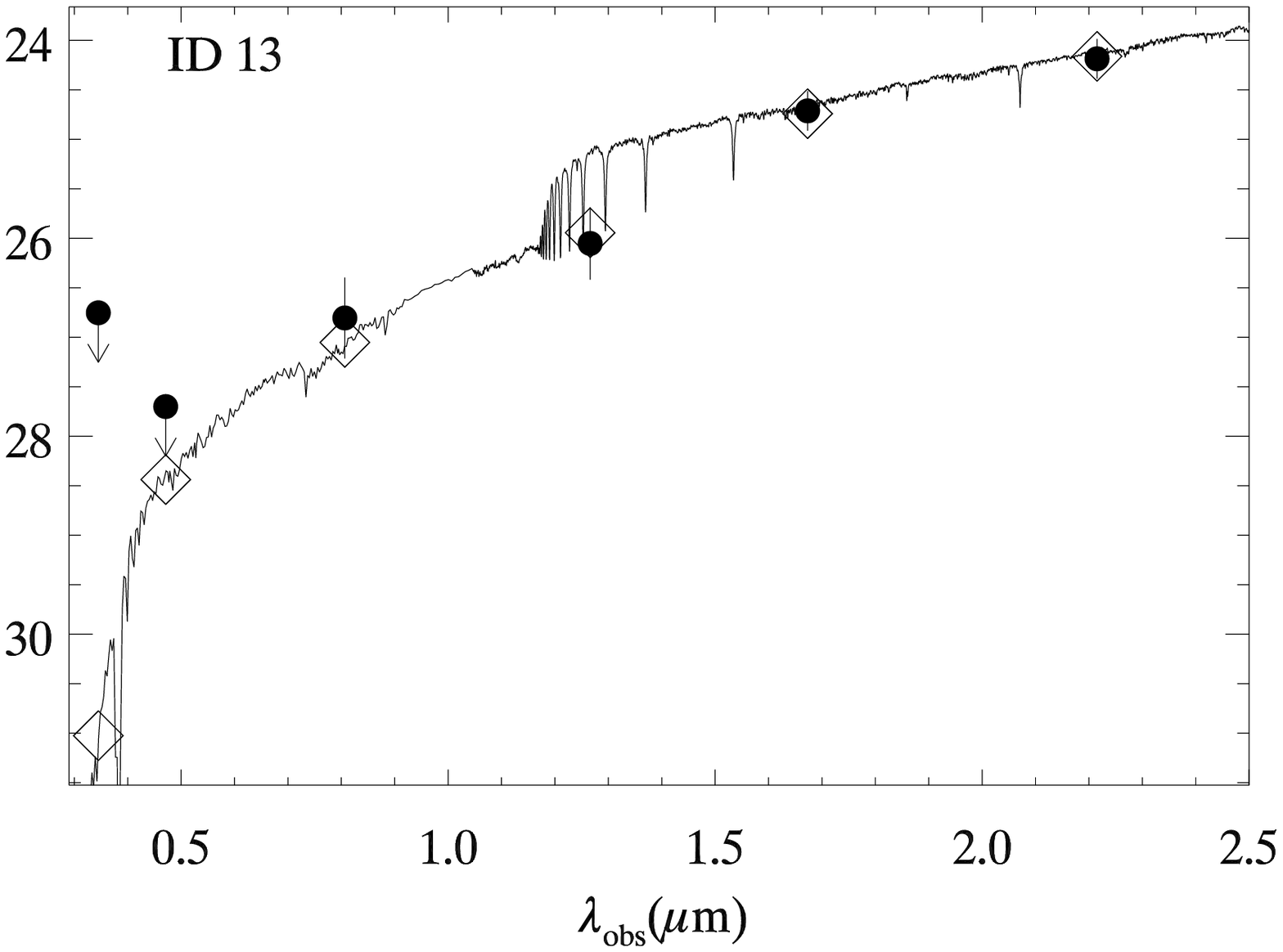}
\includegraphics[width=0.5\columnwidth]{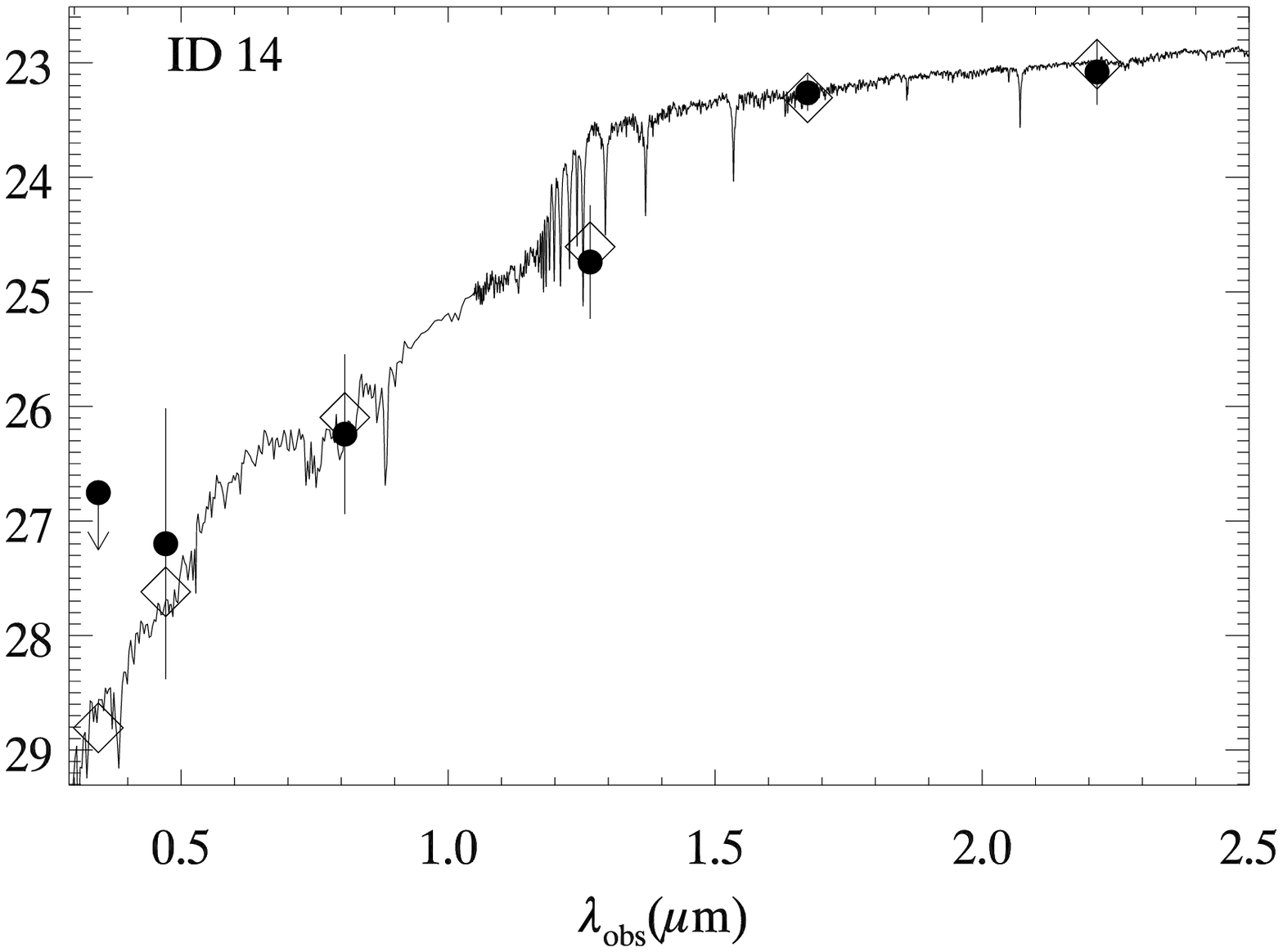}
\includegraphics[width=0.5\columnwidth]{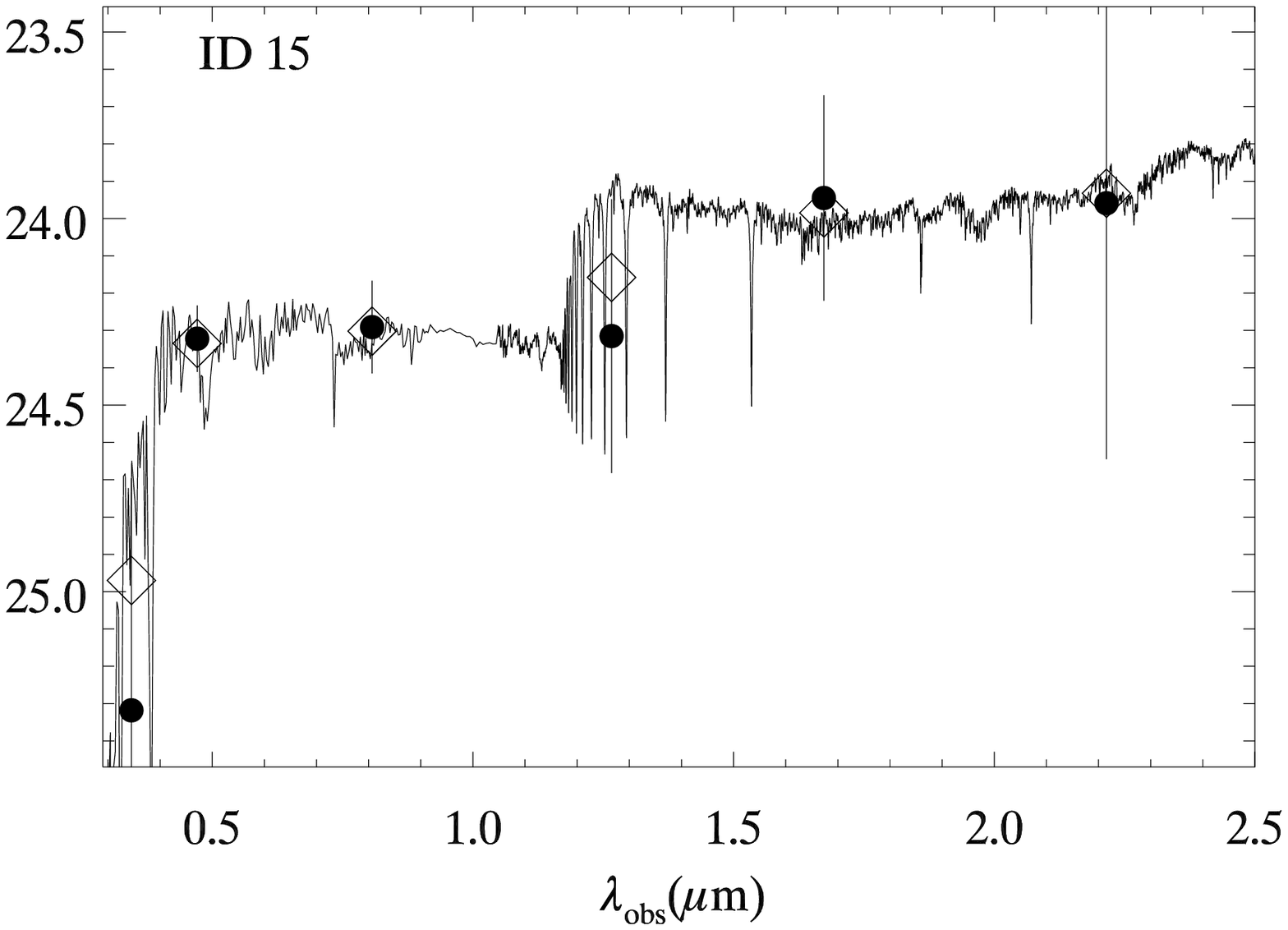}
\includegraphics[width=0.5\columnwidth]{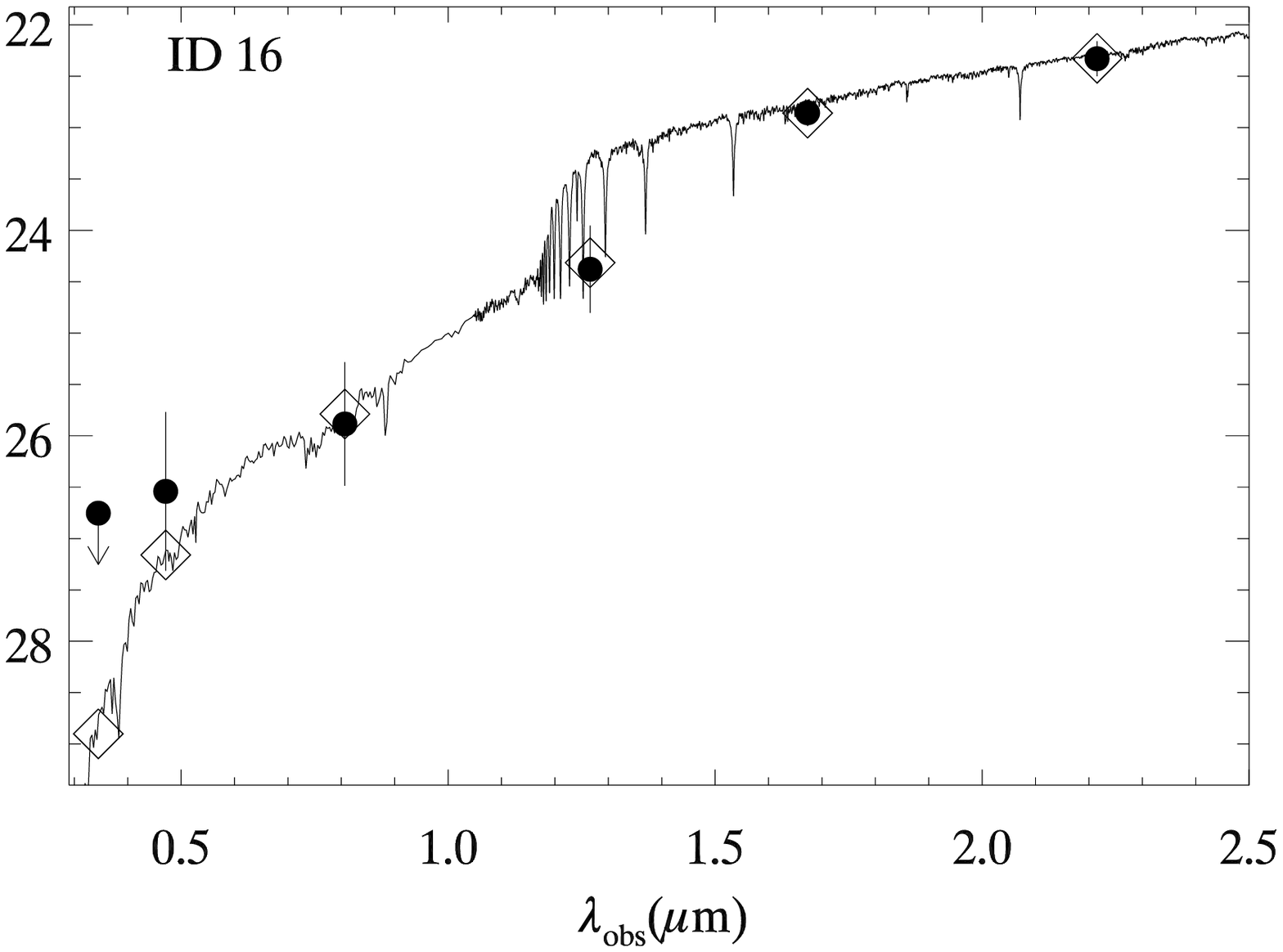}
\includegraphics[width=0.5\columnwidth]{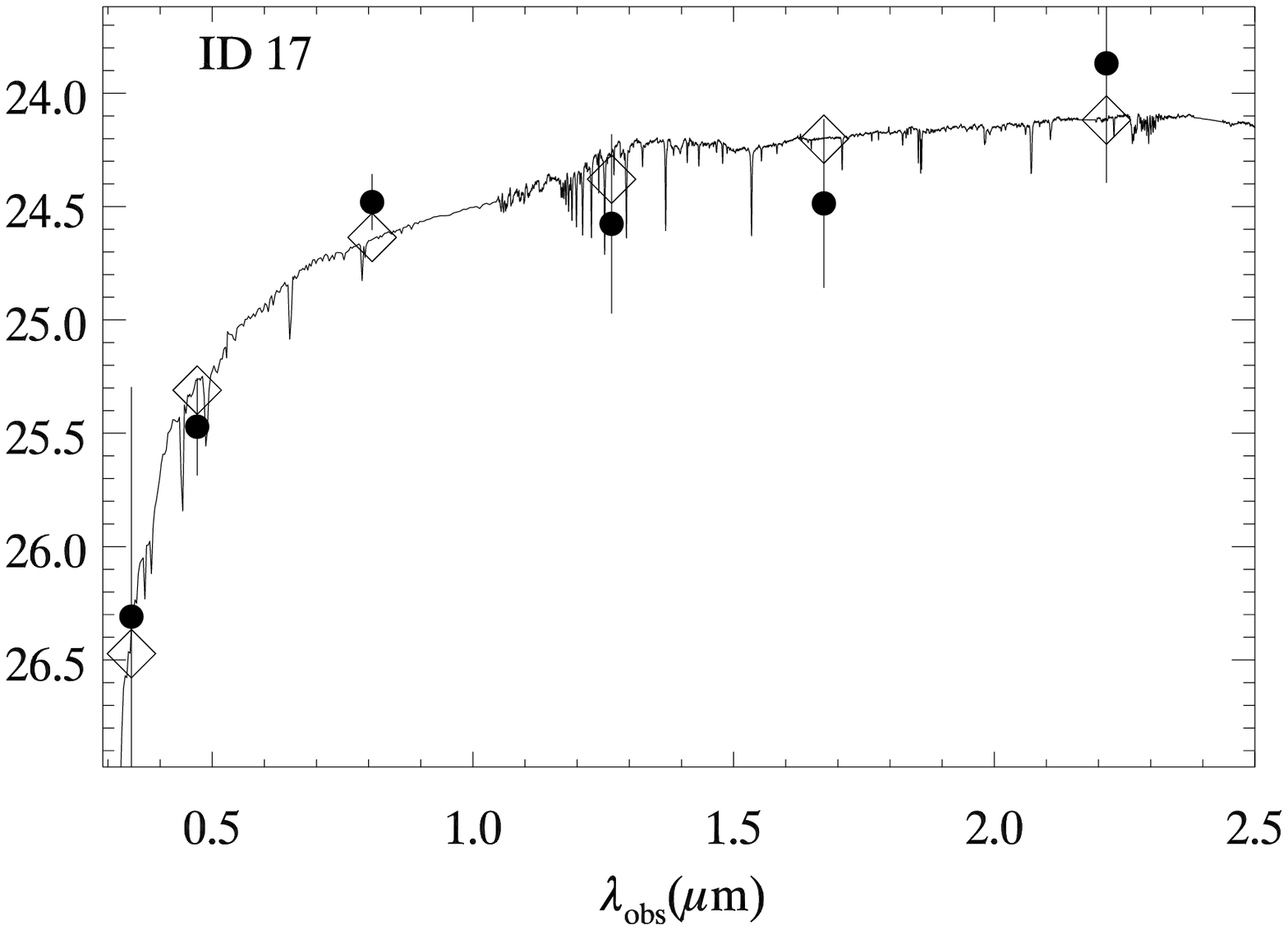}
\includegraphics[width=0.5\columnwidth]{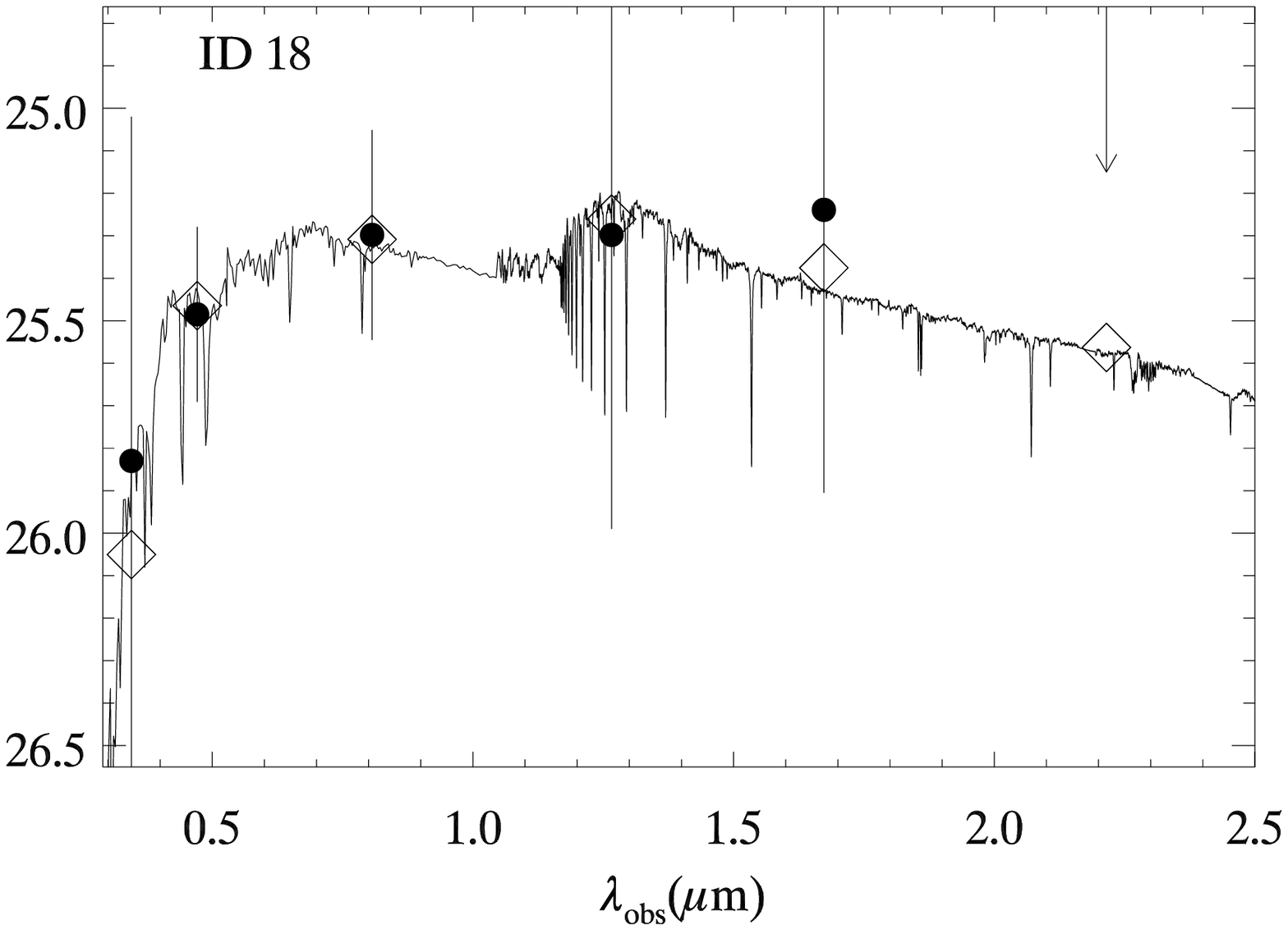}
\includegraphics[width=0.5\columnwidth]{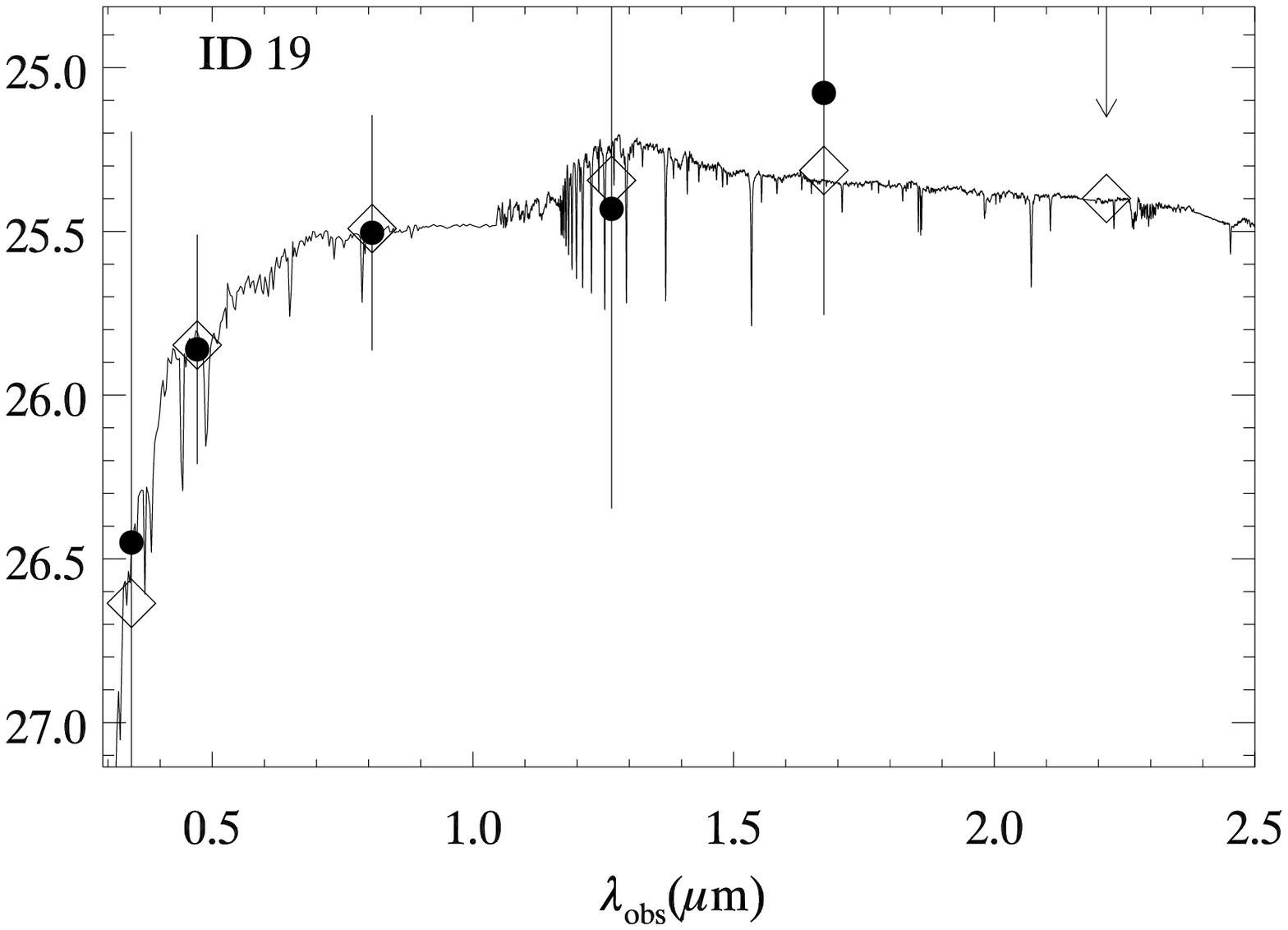}

\caption{Broad-band photometry (black filled circles) and 1$\sigma$ errors are shown for the radio galaxy (ID 1) and 18 satellite galaxies. The point source contamination from the AGN has been removed from the radio galaxy. The best-fit stellar population models are shown in grey, the diamond symbols are the convolution of the best-fit stellar population models with the filter transmission curves.  \label{SED_fits}}
\end{figure*}

\begin{table}

\begin{tabular}{lrrrr}\hline  
ID&SFR&Mass&Mass (upper limit) &Detection\\
&  (\Msunpyr)& ($10^{9}$\Msun)& ($10^{9}$\Msun)&method\\
\hline
1&
$   15.6\pm    0.7$&
$ 1100\pm200$&
$ 9900^{+  300}_{- 1300}$&
\lya,\ha, BB \\
2&
$    0.0\pm    0.4$&
$   12^{+   13}_{-    7}$&
$   16^{+   15}_{-    1}$&
BB \\
3&
$    2.2\pm    0.5$&
$   11\pm8$&
$  105^{+   18}_{-   54}$&
Photo-z \\
4&
$    0.0\pm    0.4$&
$   31^{+   24}_{-   16}$&
$   34^{+   18}_{-    8}$&
BB \\
5&
$    5.2\pm    0.5$&
$    7.2^{+    0.4}_{-    1.4}$&
$   15\pm3$&
\lya\ \\
6&
$    4.4\pm    0.4$&
$    9.3^{+    5.4}_{-    3.9}$&
$   37^{+  175}_{-    6}$&
\lya\ \\
7&
$    3.8\pm    0.4$&
$    3.4^{+    0.6}_{-    0.6}$&
$    6.9^{+    2.6}_{-    1.9}$&
\lya\ \\
8&
$    1.2\pm    0.5$&
$    5.6^{+    4.5}_{-    4.1}$&
$   72^{+   3}_{-   64}$&
\ha\ \\
9&
$    7.2\pm    0.5$&
$    5.2^{+    1.5}_{-    3.4}$&
$    9.7^{+   64}_{-    2.8}$&
\lya\ \\
10&
$    4.4\pm    0.4$&
$    0.4^{+    0.3}_{-    0.1}$&
$    0.4^{+    0.8}_{-    0.1}$&
\lya,\ha \\
11&
$    5.8\pm    0.4$&
$    2.0^{+    1.1}_{-    1.0}$&
$   38^{+   13}_{-   32}$&
\lya,\ha \\
12&
$    2.2\pm    0.6$&
$   29^{+   29}_{-    9}$&
$  111^{+  150}_{-   42}$&
\lya,\ha, BB \\
13&
$    0.0\pm    0.4$&
$    6.4^{+    6.2}_{-    2.9}$&
$   22\pm9$&
 BB \\
14&
$    0.4\pm    0.7$&
$   16^{+   18}_{-    5}$&
$   55^{+   13}_{-    10}$&
 BB \\
15&
$    5.2\pm    0.4$&
$    1.3^{+    1.0}_{-    0.5}$&
$   28^{+   18}_{-    25}$&
\lya\ \\
16&
$    0.7\pm    0.7$&
$   50^{+   35}_{-   20}$&
$  110^{+   54}_{-   31}$&
 BB \\
17&
$    1.8\pm    0.4$&
$    2.1^{+    0.8}_{-    0.7}$&
$    6^{+    4}_{-    2}$&
Morphology \\
18&
$    1.8\pm    0.4$&
$    0.2^{+    1.0}_{-    0.1}$&
$    9^{+   8}_{-    9}$&
Morphology \\
19&
$    1.3\pm    0.5$&
$    0.2^{+    1.7}_{-   0.1}$&
$    0.7^{+   20}_{-    0.2}$&
Morphology \\
\hline 
\end{tabular} 
\caption{The mass is derived from fitting the photometry to a single exponentially declining star formation history. The mass upper limit is derived from a two-model fit to the photometry, in which one model is maximally old, i.e., 3\,Gyrs and the other is maximally young (1\,Myr).  Column 5 lists the detection methods by which the galaxy was selected to be in the protocluster. \lya, \ha\ are objects which have an excess of line-emission placing them at the same redshift as the radio galaxy, BB indicates  galaxies with strong Balmer breaks inferred from large observed \jband-\hband\ colours. \label{results}}
\end{table}

\subsubsection{Stellar mass}  
The nucleus of the radio emission is associated with galaxy 1 \citep{Pentericci1997}.  The total stellar mass of this galaxy obtained from a fit to a single stellar population model is 1.1$\pm0.2\times10^{12}$\Msun\ which is consistent with the upper limit obtained from the rest-frame near-infrared light \citep{Seymour2007}. This galaxy dominates the central 150\,kpc in terms of mass, comprising 85$\pm{5}$ per cent of the total mass. If one assumes that the mass of the satellite galaxies is closer to the upper mass limit (listed in column 3 of Table \ref{results}) the fraction of mass in the radio galaxy reduced to 60$\pm13$ per cent. Rest-frame UV and optical light may be scattered from the central active nucleus contaminating the stellar light of galaxy 1, resulting in an overestimate of the stellar mass. Polarization measurements are essential to determine the fraction of scattered quasar light within each band.  The upper mass limit of galaxy 1 is an order of magnitude larger than previous measurements. We therefore always assume in the following discussion, that the mass of this radio galaxy is that of the lower mass limit ($1\times10^{12}$\Msun) which is consistent with the measurement of  \citet{Seymour2007}.

Galaxy 1 is at least $\sim$10 times more massive than any other galaxy within 150\,kpc.  The sample we have selected is unlikely to be a complete inventory of satellite galaxies. From a fit of the observed $Ks$ mag and the derived upper mass limit, we find that the 3$\sigma$ $Ks-$band detection limit of 24.4\,mag corresponds to a mass of $\sim6\times10^{9}$\Msun. Our sample is unlikely to be missing any satellite galaxies more massive than this mass limit. We identify galaxy 1 as the progenitor of the BCG, which already dominates the mass budget of the protocluster core, and we refer to this galaxy as the radio galaxy or the central galaxy. The remaining protocluster galaxies within 150\,kpc of the radio galaxy are referred to as the satellite galaxies. The radio and the satellite galaxies are collectively referred to as the Spiderweb system.

All galaxies have sufficient red light to be habouring an older stellar population. The maximum masses of the galaxies are generally 1--20 times greater than the mass derived from fitting a single exponentially declining stellar population. The exception is galaxy 18 whose SED allows up to 45 times more mass to be hidden beneath the young stellar population that dominates the light in the observed UV and optical wavelength range.

\subsubsection{Star formation rates}
The total star formation rate of all the galaxies determined from the UV emission is 63$\pm8$\Msunpyr. The star formation is predominately located within the satellite galaxies rather than in the central galaxy. The radio galaxy contains $\sim25\pm2$\% of the star formation occurring within the whole system. Dust extinction in the radio galaxy may skew the fraction of UV light observed in the radio galaxy compared to the satellite galaxies.  This percentage does not take into account the diffuse star formation that lies between the radio and satellite galaxies. \citet{Hatch2008} find that the amount of intracluster star formation is similar to the amount of star formation that occurs within the galaxies in a similar sized area (15\,x\,18\,arcsec$^2$). Thus most of the star formation occurs beyond the region we have defined as the central galaxy. Only 3 satellite galaxies have no detected Far-UV flux, the majority are in a phase of rapid star formation. From the current data we cannot determine whether galaxies 2, 4, and 13 are passively evolving red galaxies, or dusty star-forming galaxies.

Large amounts of dust-obscured star formation is confirmed by the bright sub-millimeter (sub-mm) emission observed in this galaxy \citep{Stevens2003}, with a 850$\micron$ flux density of 6.7$\pm{2.4}$\,mJy. Thus the total star formation rate of the region close to the radio galaxy is potentially much higher than given by the dust-uncorrected UV luminosity. The galaxies have red UV colours (see Fig.\,5 in section \ref{sec:colour}) implying extinctions in the approximate range $0.1<E$($B-V$)$<0.5$ depending on the age, star formation history and metallicity of the stellar population. Correcting the UV luminosity for this amount of dust extinction would bring the star formation rates derived from the observed UV and sub-mm emission into agreement. The 850$\micron$ emission from this location is spatially elongated along a position angle of 72$^{\circ}$ \citep{Stevens2003}. The satellite galaxies with extreme star formation and large amounts of extinction (galaxies 5, 6, 7, 9, 10, 11 and 15) are elongated approximately East-West along the sky stretching 15\,arcsec from end-to-end. Therefore the extended sub-mm emission results from source confusion and is the sum of the combined star formation occurring in the radio and satellite galaxies. High spatial resolution sub-mm observations have deconvolved the extended sub-mm emission from the high redshift galaxy 4c40.7 into two separate components indicating merger induced star formation \citep{Ivison2008}.  These optical and near-infrared observations show the star formation occurs in multiple compact objects and high resolution sub-mm may be able separate the sub-mm emission into the various galaxy components.

\subsubsection{Spatial distribution of the satellite galaxies}
Almost all of the satellite galaxies lie within the large \lya\ halo that surrounds the radio galaxy, even those galaxies with little or no star formation. Since our observations show a single snapshot in time, most of the satellite galaxies' orbits must lie within the \lya\ halo. This gives an approximate orbit radius of 100\,kpc. The asymmetric \lya\ halo extension to the East also contains satellite galaxies. This strong spatial correlation implies a connection between the \lya\ halo and the satellite galaxies. The \lya\ halo may originate from gas stripped from the satellites, or the galaxies may excite the gas, possibly by shocks from their movement through an existing gas reservoir, or UV emission from hot young stars. The total \lya\ luminosity of the halo is 2.6$\times10^{45}$\ergps\, and would require a star formation rate of $\sim$2500\Msunpyr\ to ionize it. The UV star formation rates are more than an order of magnitude too small to ionize the halo, although the bright sub-mm emission suggests a large amount of obscured star formation. It is possible that the \lya\ halo is ionized by the UV emission from young hot stars produced during this phase of extreme star formation, or is a mixture of both stellar and AGN photoionization \citep{VillarMartin1997,VillarMartin2007}.

\subsection{Comparison of the Spiderweb system to simulations}
\label{simulations}
\begin{figure}
\centering
\includegraphics[width=1.0\columnwidth]{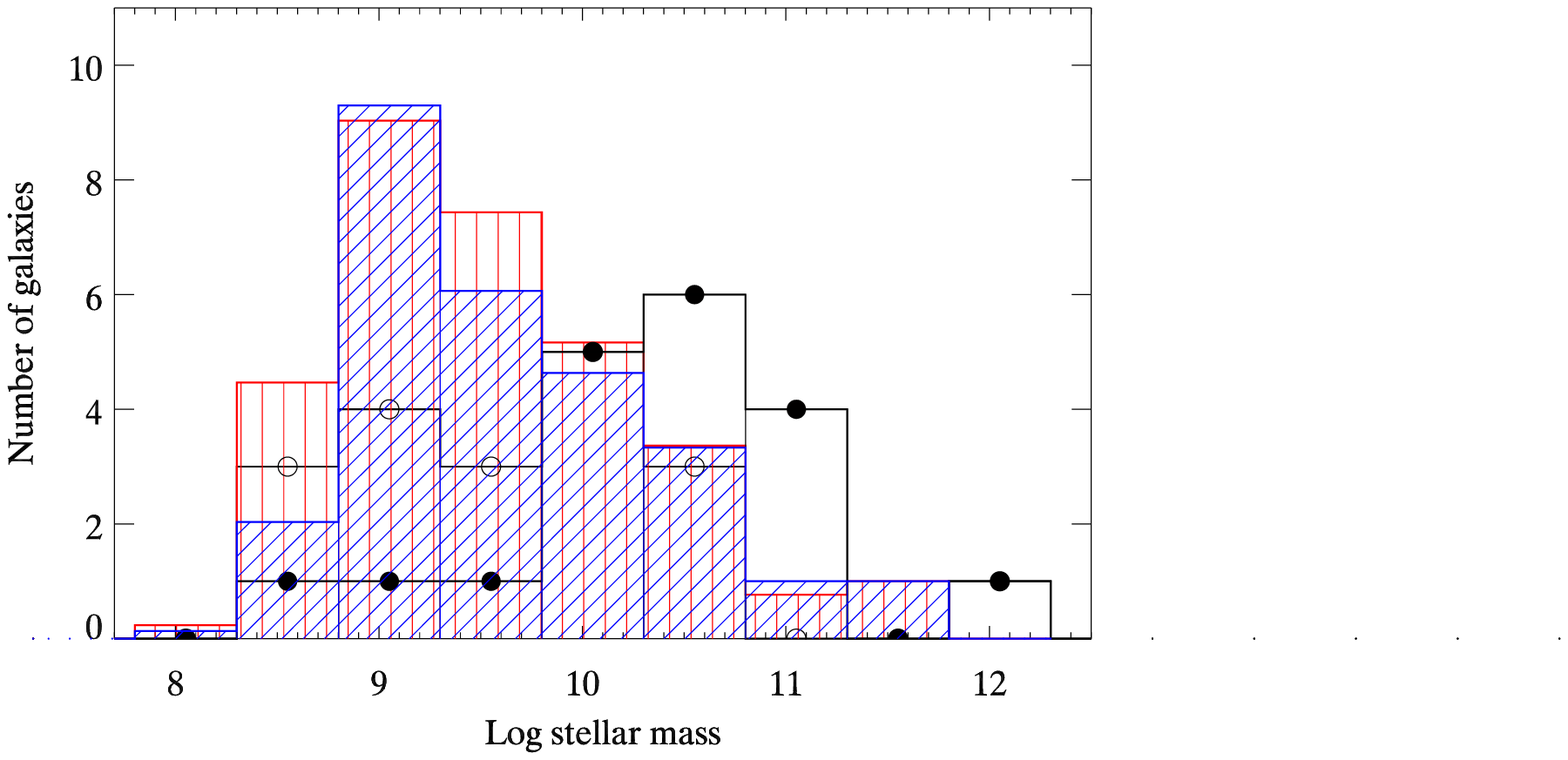}\caption{Comparison between the derived stellar masses of the observed galaxies within 150\,kpc of the Spiderweb galaxy (black histograms) and the masses of galaxies within the same projected distance from the 10 most massive galaxies within  2 SAMs (F08 -- blue, and DBL07 -- red). Poisson errors for the SAMs are given by $\sim$(number in each bin)$^{1/2}\times0.18$. Only 
those simulated galaxies which have $M_{H}$ less that the $3\sigma$ detection limit (26.6 mag) are plotted. Derived upper mass limits are plotted as black solid circles, and lower mass limits for the satellites are plotted as open circles.  \label{mass}}
\end{figure}

We compare our results to two different semi analytic models (SAMs): \citealt{deLucia2007} (DLB07) and \citealt{Font2008} (F08) which are based on the dark mater only Millennium simulation\footnote{{\rm http://www.mpa-garching.mpg.de/millennium/}}.   We construct catalogues from the SAMs that most resemble our own selection criteria.  We select all galaxies within 10\,Mpc of the most massive (stellar mass) 10 galaxies at $z=2.2$, then make a further cut selecting only those galaxies that lie within a projected  distance of 150\,kpc from the massive galaxy. We take an average of 3 orthogonal projections of each of the 10 galaxies.

We compile catalogues of the SDSS\,$g'$, SDSS\,$i'$, $H$ and $K$ magnitudes for all simulated galaxies. Although the filter curves differ slightly to the observed \gband, \iband, \hband\ and $K_{s}$ filters, this will not affect our comparisons as we only consider large (1\,mag) bins when comparing galaxy magnitudes. We convert the SDSS\,$g'$ luminosities to star formation rates using equation \ref{sfrg} for comparison to the observed UV star formation rates.  The SAMs predict a large number of low luminosity satellite galaxies that may bias any comparison to the data, therefore we make a luminosity cut equivalent to the 3$\sigma$ detection limit in the \hband. 

The average number of galaxies within each projected box is 27 for the F08 models and 30 for DBL07 compared to the 19 galaxies observed within 150\,kpc of MRC\,1138-262. The SAMs predict too many satellite galaxies above our detection threshold compared to our observations. This discrepancy is further compounded by the fact that the dark matter halo of MRC\,1138-262 is probably more massive than its simulated counterparts in the SAMs. The missing simulated satellites are faint and have low masses (see Fig.\,\ref{mass}). Such faint, low-mass galaxies may not be resolved in the smoothed images, especially if they are not compact. A larger number of galaxies are seen in the higher resolution {\it HST} \iband\ images of \citet{Miley2006} of the same region, and the intracluster light discussed in \citet{Hatch2008} may be partly due to unresolved, low mass satellites. It is therefore likely that the observations presented here miss some low-mass satellite galaxies because they cannot be resolved, and thus the discrepancy is most likely due to missing satellite galaxies in the observations, rather than the simulations predicting too many satellite galaxies.
 
\subsubsection{Stellar Masses}
The stellar mass of the most massive semi analytic galaxy at $=2.2$ is $2.2\times10^{11}$\Msun\ in the F08 models and $2.8\times10^{11}$\Msun\ in the DLB07 models. These are $\sim$4 times less massive than the radio galaxy (galaxy 1). Our derived mass of $10^{12}$\Msun\  for the radio galaxy places it as one of the most massive galaxies detected at $z>2$. However, we note that this mass is likely to be an upper limit, as we have not taken into account light scattered from the central AGN. Mass derived from fitting stellar templates to broadband photometry are systematically overestimated by a factor of $1.4-2$ compared to masses derived from fitting stellar templates to spectroscopy \citep{Shapley2005,Kriek2008}. We must also consider that stellar population synthesis modeling contains large systematic uncertainties and the mass of these galaxies, especially galaxy 1, may be lower than the presented in Table \ref{results}. Most high redshift radio galaxies at $z\sim2$ have similar masses to the most massive galaxies in the SAMs \citep{Seymour2007}, the Spiderweb Galaxy has an exceptionally high mass. 

In general the central galaxy in each model contains between 5 and 35 per cent of the stellar mass within 150\,kpc of the massive galaxy. If we assume the upper mass limits for the satellite galaxies, the central galaxy in the Spiderweb system contains 60$\pm13$ per cent of the stellar mass. Although the stellar mass distribution of the SAMs is generally not as centrally concentrated  as the Spiderweb system, one of the massive galaxies in the F08 SAM contains 35 per cent of the selected stellar mass within the central galaxy, and is therefore compatible with the lower limit allowed by observations. Thus the Spiderweb system may not typical,  but is compatible with of the stellar mass distribution of massive galaxies in these SAMs.

The SAMs of \citealt{Saro2008a} based on a dark matter plus non-radiative gas particles produce BCGs that are 25 per cent larger than those from dark-matter only simulations, and their dominant BCGs contains twice as much stellar mass as any satellite galaxy. These simulations appear to match the derived mass distribution of the Spiderweb system better than the SAMs based on dark-matter only.

Fig.\,\ref{mass} compares the derived upper and lower limits of the stellar masses of the galaxies within 150\,kpc of the radio galaxy to those galaxies within 150\,kpc of the 10 most massive galaxies in the SAMs at $z=2.2$. Filled circles plot the upper limit masses, whilst open circles plot the lower limit masses. We only plot simulated galaxies with $m_H<26.6$, the 3$\sigma$ detection limit in \hband. 

The stellar masses of the simulated central galaxies are at least a factor of 4 less that the observed radio galaxy, therefore we are comparing the observations to the masses of galaxies in lower mass halos that of the Spiderweb Galaxy. We thus expect the true distribution  of galaxy masses to be shifted to higher masses compared to these simulations. Above $10^{10}$ \Msun, the SAMs agree with the derived lower mass limits, whilst the upper mass limits lie to the right of the simulated galaxy mass distribution. This is entirely consistent with our expectations and imply that the derived upper and lower mass limits are likely to bracket the true stellar mass of the galaxies. 
Below  $10^{10}$\Msun\ the simulations predict a larger number of satellites  than are observed. Comparing our upper mass limits to the observed $Ks$ magnitude,  we find we are only complete to masses above $\sim10^{9.8}$\Msun, thus we are likely to be missing many of the low-mass satellites with red colours.

\begin{figure*}
\centering{
\includegraphics[width=2.0\columnwidth]{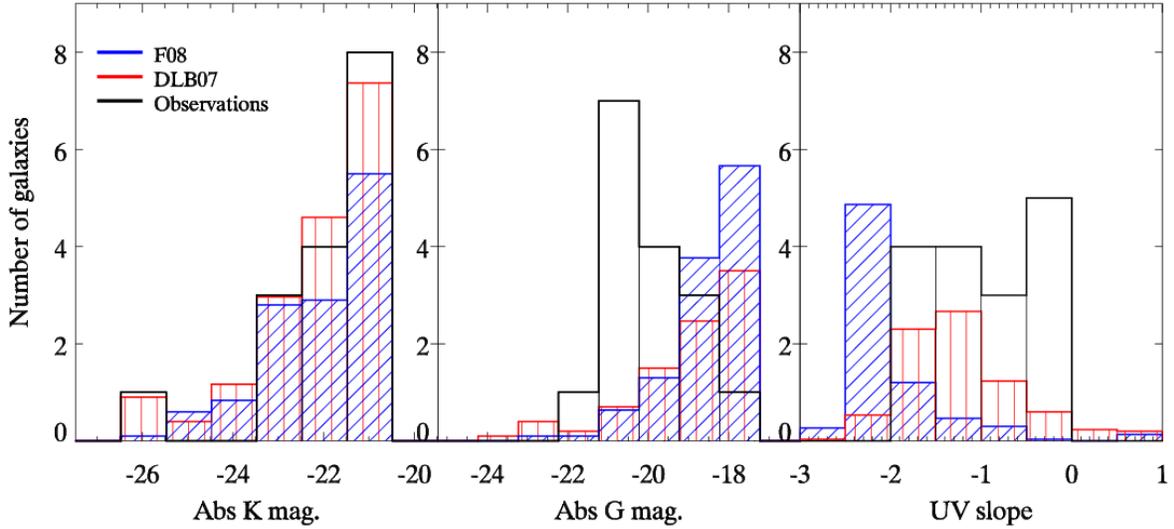}\caption{$K$ and $G$ band luminosity functions and UV slope for the observed galaxies (black) and  2 versions of SAMs (F08 -- blue, and DBL07 -- red). Poisson errors for the SAMs are given by $\sim$(number in each bin)$^{1/2}\times0.18$. Both luminosity functions stop at the 3$\sigma$ detection limit in the considered band. The observed $K$ luminosity function matches reasonably well with the models, suggesting that the mass of the simulated galaxies matches the observations. However there is a mismatch in the observed \gband\ and $G$ luminosity functions implying the star formation rates in the simulated galaxies are too low. Both models produce galaxy UV slopes that are too blue, although the DBL07 models are a better match to the observations. This implies the simulated galaxies which have active star formation have either too little dust or the stellar ages are too low. \label{LF}}}
\end{figure*}

\subsubsection{Galaxy colours}
\label{sec:colour}
The right-most panel of Fig.\,\ref{LF} shows a comparison of the range of UV continuum slope, $\beta$, for the observed and simulated galaxies within 150\,kpc of the radio galaxy. The simulated galaxies have been subjected to the same select criteria as the observations -- the galaxies must be observed in both the $i'$ and $g'$ bands -- thus the galaxies must be actively forming stars.  The red UV colours suggest significant amounts of dust between us and the galaxies. The amount of extinction, $E$($B-V$), that can cause these red colours can range from 0.1 to 0.5 depending on the age and metallicity of the galaxy's stellar population. DLB07 models provide the closest fit to the data, however, the simulated galaxies with active star formation have too little dust or the stellar ages are too low. 

\subsubsection{Star formation rates}

The SAMs predict total star formation rates of 30--70\Msunpyr\ within 150\,kpc radius, similar to the observed UV star formation rate of 63$\pm8$\Msunpyr\ (dust-uncorrected) in the Spiderweb system. The galaxy colours show that the dust content of the simulated galaxies is less than that of the observed galaxies, and therefore a comparison of the intrinsic star formation rates may result in a discrepancy.
  
We make a more detailed comparison by comparing the observed $K$ and  $g$ magnitudes with those from the SAMs in Fig.\,\ref{LF}. The left panel shows there is general agreement between the observed $Ks$ luminosity function and the SAMs. The $Ks$ band is the reddest observational band and covers rest-frame $V-$band light. From the  selection of observational data used in this work, it is the best tracer of galaxy mass, however there can be significant contamination of light from young stars. 

The $g-$band covers rest-frame 1500\AA\ light, which can be converted to star formation rate through equation \ref{sfrg}. This comparison is shown in the middle panel of Fig.\,\ref{LF}. We note that the observed \gband\ luminosity function does not have the usual shape of a UV luminosity function.  Therefore we do not expect the simulations, which reproduce the general galaxy population at $z\sim2$ \citep{Guo2008}, to match the observed rest-frame UV luminosity function. Indeed the general shape of the simulated UV luminosity function agrees with studies of the $z\sim2.2$ luminosity function (e.g., \citealt{Sawicki2006}). However this means the simulated satellite galaxies typically have lower $g$ magnitudes, and therefore lower star formation rates, than the observed satellite galaxies in the Spiderweb system.  Since the number of simulated galaxies at a given mass approximately matches the observations, this discrepancy is caused by the low star formation rates of the simulated galaxies, rather than a lack of satellite galaxies. The SAMs predict centrally star formation, whilst the observations show that most of the star formation occurs beyond the central radio galaxy. The SAMs reproduce the general galaxy population at $z\sim2$, but are unable to match some of the details of this special system.

\subsubsection{Smooth-particle hydrodynamical simulations}
\citet{Saro2008b} simulate two protoclusters using smooth particle hydrodynamical simulations and have compared their simulations to observations of the Spiderweb Galaxy, and the galaxies within 75\,kpc of this radio galaxy. The BCGs from these smooth-particle hydrodynamical simulations have stellar masses of $5\times10^{12}h^{-1}$\Msun\ and $8\times10^{11}h^{-1}$\Msun at $z=2$. However, these simulations do not include AGN feedback, which will act to reduce these masses, thus these masses should be considered upper limits. The most massive satellite galaxies within the simulations are a factor of 4-10 less massive that the already dominant BCG. Therefore the stellar mass distributions of the smooth-particle hydrodynamical simulations are different to that of the SAMs, and they provide a better match to the observed stellar mass distribution of the Spiderweb Galaxy.

The smooth particle hydrodynamical  simulations have total star formations rates of 600\Msunpyr\ and 1750\Msunpyr within a smaller region (within a box of 150\,h$^{-1}$kpc on a side) than that considered in the data presented here. These high rates are much greater than observed in the UV, although the bright sub-mm emission from this source \citep{Stevens2003} indicates a large amount of obscured star formation of the order of a thousand solar masses per year. Thus the \citet{Saro2008b} results are consistent with the data.

A comparison to the luminosities of the  simulated galaxies of \citet{Saro2008b} is not possible, but we note that they match the \iband\ galaxy luminosity function of the galaxies within 75\,kpc of MRC\,1138-262 once they include a prescription for dust attenuation.

\section{Discussion}
\label{discussion}
\subsection{The state of play at $z=2.2$}
Within a 150\,kpc radius, the high redshift radio galaxy MRC1138-262 at $z\sim2.2$ is surrounded by 18 satellite galaxies with masses between $10^{8}-10^{11}$\Msun. Most of the  galaxies within 150\,kpc of the radio galaxy are UV-bright line-emitters and have large star formation rates.  These star forming galaxies have similar upper limit masses to the galaxies selected through their strong Balmer breaks. Thus the galaxies with sufficient gas reservoirs to fuel large star formation rates do not necessarily have low stellar masses.  It is known that star formation can be triggered by merging (e.g.\,\citealt{Mihos1994}) and it is likely that this is occurring on a large scale in this system, with most ($>75\pm2$ per cent) of the stars forming in the satellite galaxies, rather than in the central galaxy. 

With a mass of $10^{12}$\Msun\ the central galaxy is already of a similar mass to cD galaxies at the centre of nearby galaxy clusters.  The fraction of mass in the Spiderweb system that is contained within the satellite galaxies, depends on whether we assume the stellar masses are better estimated by the  lower or upper mass limits, given in columns 3 and 4 of Table \ref{results} respectively. Assuming upper (lower) mass limits, the combined mass within all the satellite galaxies comprises $40\pm13$ ($15\pm5$) per cent of the total stellar mass.  The radio galaxy still has room for further growth, and may undergo a number of major-mergers before it reaches the limit of the most massive, local cD galaxies.  However, galaxy 1 has a $Ks-$band magnitude that is 2 magnitudes brighter than any other protocluster galaxy candidate known within a 1.5\,Mpc radius \citep{Kurk2004}, and therefore it is unlikely that the protocluster contains more galaxies of a similar mass at the observed redshift.

\citet{deLucia2007} find that $\sim70$ per cent of BCGs at $z=2$ do not become BCGs at $z=0$, but rather end up as the central galaxies of massive halos which are less massive than cluster halos. The most massive halos at $z=2$ are generally not the progenitors of $z=0$ clusters. Therefore, whilst MRC\,1138-262 appears to the the brightest protocluster galaxy at $z=2$, it is uncertain whether the Spiderweb system will evolve  into a local BCG.

\subsection{The evolution of the Spiderweb System}
\subsubsection{Merging timescales}
A complete N-body numerical investigation of this system is required to study the evolution of the Spiderweb system, which comprises of the radio galaxy and all galaxies within a 150\,kpc projected radius. The data given in Table \ref{results} and Fig.\,\ref{overlay} can be taken as initial conditions for such a simulation. However, if we assume that the satellite galaxies all lie on circular orbits around the radio galaxy, with radii given by their projected radii, we can use analytic approximations to gain an idea of the evolution of this system.
 
When an orbital point mass (such as the satellite galaxy) travels through an uniform background mass distribution (such as the dark matter halo of the central galaxy), it experiences a steady deceleration known as dynamical friction \citep{Chandrasekher1943}. The merging timescale ($\tau_{\rm merge}$) of these galaxies can be taken as the dynamical friction timescale given by
%or streaming 0808.0553
\begin{equation}
\tau_{merge}=1.17\frac{\tau_{dyn}}{\rm ln \Lambda}\left(\frac{M_{\rm central}}{M_{\rm satellite}}\right)
\label{df1}
\end{equation}
where the Coulomb logarithm ${\rm ln \Lambda}={\rm ln}(1+ M_{\rm central}/M_{\rm satellite})$, $M_{\rm central}$ is the mass of the central galaxy, $M_{\rm satellite}$ is the mass of the satellite galaxies and the orbital timescale is given by $\tau_{dyn}=(\frac{R^3}{GM_{\rm Total}})^{1/2}$, where $R$ is the distance between the satellite and central galaxy \citep{Binney1987}. $M_{\rm central}/M_{\rm satellite}$ is the ratio of dark matter halo mass. Matching the abundances of galaxies above a given stellar mass to halos above a given dark-matter mass is a powerful technique employed to relate the galactic stellar mass to a halo mass  \citep{Conroy2008b}, however, this technique is not reliable at high redshift ($z>1$).  Instead we assume that the amount of dark matter within each galaxy is directly proportional to the galaxy's stellar mass, therefore the ratio $M_{\rm central}/M_{\rm satellite}$ is the ratio of stellar mass.  If the abundance-matching estimates of \citet{Conroy2008b} are correct, the merging timescales will be reduced, especially for low mass satellites. $M_{\rm satellite}$ is the mass of the satellite galaxies and $M_{\rm central}=1\times10^{12}$\Msun, the lower limit of the stellar mass of galaxy 1. The total (dark matter plus stellar) mass of the central galaxy is estimated from the available dynamical information. Spectroscopic observations of galaxies 5, 7, 9, 11, 15 show they have line-of-sight velocities of 1030\kmps, 210\kmps, 570\kmps,1180\kmps\ and 1240\kmps\ relative to the central galaxy \citep{Miley2006} which are consistent with a central mass of $M_{\rm Total}=10^{13}$\Msun.

\citet{BoylanKolchin2008} used numerical simulations to find a more realistic merging timescale for minor mergers. The merging timescale is increased relative to equation \ref{df1} due to the significant mass loss of the satellites as they orbit the central galaxy. The merging timescale is given by 
\begin{equation}
\tau_{merge}=0.216\frac{\tau_{dyn}}{\rm ln \Lambda}\left(\frac{M_{\rm central}}{M_{\rm satellite}}\right)^{1.3}{\rm exp}[1.9(1-e^2)^{1/2}], 
\label{df2}
\end{equation}
where $e$  is the orbital eccentricity. We assume all satellites have circular orbits ($e=0$). Lower angular momentum orbits have shorter merging times, if the satellites have elliptical orbits ($0<e<1$), their merging times could decrease by a factor of $\sim7$.

\begin{figure}
\centering
\hspace{-15mm}
\includegraphics[width=1.15\columnwidth]{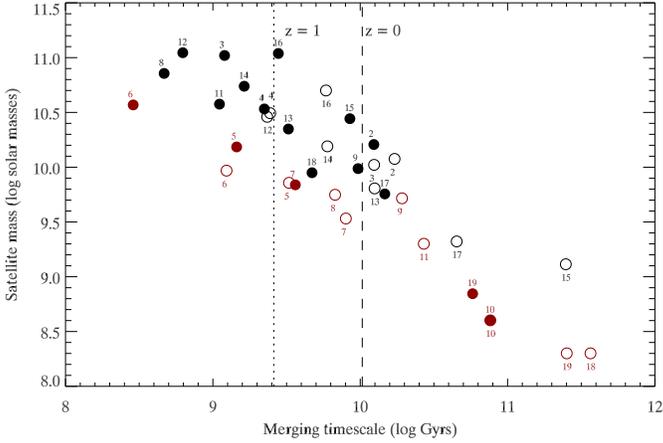}
\caption{The merging times for all satellite galaxies within 150\,kpc of the central radio galaxy. The dotted and dashed lines indicate the time from $z=2.2$ until $z=1$ and $z=0$ respectively. Merging timescales are derived from equation \ref{df2}. Filled circles indicate merging times using the upper mass limits from column 4 of Table \ref{results}, whilst the open circles are merging timescales derived assuming the satellites have the lower masses from column 3. Red symbols are galaxies who are likely to be experiencing rapid tidal stripping as their tidal radii are less than their half-light diameters. Most satellites will either merge with the central radio galaxy by $z=0$, or are experiencing rapid tidal stripping, and are therefore unlikely to survive until $z=0$ as separate entities.  \label{merge}}
\end{figure}

The merging times for all satellite galaxies within 150\,kpc of the central radio galaxy are shown in Fig.\,\ref{merge}. The filled circles mark the merging timescales derived using equation \ref{df2} and the upper limit of the galaxy mass, whilst the open circles use the lower limit of the galaxy mass. The central galaxy is always assumed to have a mass of $1.1\times10^{12}$\,\Msun, which is consistent with other measurements from the rest-frame near-infrared. The open circles indicate the longest possible merging times, as we have assumed both the lowest masses and orbital eccentricities of the satellite galaxies. The vertical lines show the time from $z=2.2$ until $z=1$ and $z=0$.

The more massive galaxies generally merge earliest, irrespective of their initial radial separation from the central radio galaxy.   Most of the satellite galaxies will merge by $z=0$ if they have masses close to their upper mass limits. In doing so the radio galaxy will increase its mass by a factor of 1.5$^{+0.4}_{-0.3}$. Alternatively, the mass of the satellite galaxies may be closer to the lower mass limits. In this case, only a third of the satellites will merge before $z=0$. Thus the amount of growth  is dramatically reduced, with only a $\sim$13 per cent increase in the stellar mass through merging. However, this is not the full story, as tidal stripping is an important process that also depends on the mass of the satellite galaxies.  The minimum merging timescale is $\sim300$\,Myr which allows the satellites to evolve before they merge. Thus many of them are likely to be older, redder and have lower star formation rates by the time they merge. The gas reservoirs in the satellites  may no longer exist if the satellites continue to form stars rapidly.

\subsubsection{Tidal stripping}
The typical orbital time is 100\,Myrs, therefore the satellite galaxies will orbit the central galaxy between several and hundreds of times before they merge.  As the satellites orbit the central galaxy, tidal interactions can strip large quantities of gas and stars from the satellite galaxies into tidal tails, which are only partly re-acquired by the galaxies \citep{Matteo2007}.  After several orbits kinematically distinct streams will appear. One possible such stream lies between galaxies 17 and 18 (see Figs.\,\ref{full_colour} and \ref{overlay}).

\begin{figure}
\centering
\hspace{-15mm}\includegraphics[width=1.15\columnwidth]{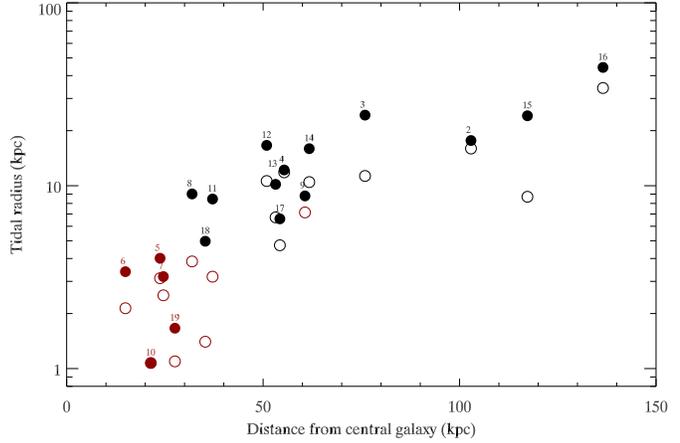}\caption{The tidal radius of the satellite galaxies derived from the upper mass limit of the satellites shown as filled circles, whilst radii derived from lower mass limits  are open circles. Red points are galaxies whose tidal radii are smaller than their half-light diameters as measured from the \hband\ image. All of these satellites are likely to be experiencing some tidal stripping, but those coloured red will be stripped at a faster rate. \label{stripped}}
\end{figure}

The tidal radius of the satellites can be approximated by 
\begin{equation}
R_{tidal}\simeq\left( \frac{M_{satellite}}{3M_{central}}\right)^{\frac{1}{3}}R~, 
\end{equation}
where $R$ is the distance between the satellite galaxy and the central radio galaxy \citep{Binney1987}. This radius approximately marks the limit within which stars are bound to the satellite, and beyond which stars can be tidally stripped. Fig.\,\ref{stripped} shows that the tidal radius for many of the satellite galaxies is less than 10\,kpc, hence the outer stars of the satellites are likely being stripped away. Further stripping may occur from interactions between the satellite galaxies, including those galaxies below the observational detection limit. The mass ratios of the satellite galaxies are closer to unity, which increases the tidal radii. However, the radial separation between two satellites may become small during some part of their orbits, and during this period the forces between the satellite galaxies may dominate, increasing the amount of tidal stripping that occurs. Such interactions can only be modeled by a N-body numerical simulation of the system.  

The red symbols in Figs.\,\ref{merge} and \ref{stripped} mark the galaxies which have smaller tidal radii than half-light diameters, measured from the \hband\ image. These galaxies are probably experiencing tidal stripping of their central stars, and it is unlikely that these galaxies will survive as distinct entities for a large number of orbits. It is likely that most of their stars will be stripped before they merge. There is a strong dependance on mass, so the satellite galaxies would experience more stripping if their mass was closer to the lower mass limit rather than the upper mass limit.

Fig.\,\ref{stripped}  shows that the stars and gas can be stripped from the satellite galaxies at large distances, more than 50\,kpc from the central galaxy.  Therefore these stripped stars may form an extended stellar halo around the radio galaxy, so the radio galaxy can grow in size, or form something akin to a cD halo. It has been postulated that the extended stellar halos of cD galaxies may form as stars are stripped during mergers \citep{Gallagher1972,Malumuth1984} and these observations may have captured the Spiderweb Galaxy in the early stages of such a process. A numerical N-body simulation of this system will  determine whether such conditions can result in an extended stellar halo such as a cD halo.

\subsubsection{Assembly scenarios of the Spiderweb system}
The  evolution of the system strongly depends on the mass ratio of the central and satellite galaxies. Since we have assumed that the central galaxy has a mass of $10^{12}$\Msun, consistent with previous measurements, the evolution of this system depends on the mass of the satellite galaxies. If the galaxies have true masses similar to the derived upper mass limits, most of the satellites will merge by $z=0$, with the most massive systems merging within a few hundred Myrs. Tidal stripping is most important for the low mass satellites that are within 30\,kpc of the central galaxy. The stars accumulated from the mergers and stripping are likely to be deposited close to the central galaxy, and the central galaxy can almost double its $z=2.2$ stellar mass through these minor mergers.

Alternatively, if the masses of the satellite galaxies are similar to the lower limits we have presented, only the third most massive satellites will merge with the central galaxy by $z=0$, and most merging occurs at later times, between $1<z<0$. The mass of the central galaxy is not greatly increased by these mergers, although this is due to the low masses of the satellites, rather than the low number of mergers. The other satellites are unlikely to survive as distinct entities until $z=0$. Fig.\,\ref{merge} shows that those satellites that will not merge by $z=0$ are experiencing strong tidal stripping of their inner stars. These  stars will be stripped from the satellites and become bound to the central galaxy. In such a  manner the central galaxy may accumulate all the stellar mass that is currently observed in the satellites. The satellites experiencing strong tidal stripping generally have the lowest mass, and longest merging timescale, and can be stripped at large separations from the central galaxy, up to 60\,kpc.

In both scenarios the central galaxy is likely to accumulate most of the stellar mass of the satellite galaxies by $z=0$, however, it does so in different manners. If the satellites' masses are large, most of the stellar mass from these satellites is deposited relatively quickly and at small radial distances from the core of the central radio galaxy. On the other hand, if the satellites have low masses, most mass is deposited through relatively slow mergers. A fraction of the stellar mass will be deposited at large radii from the core of the central galaxy by tidal stripping, forming an extended stellar halo.  The total growth of the central galaxy due to the consumption of the satellite galaxies is between $\sim$10 and 120 per cent. This large range is due to the large uncertainty of the stellar mass derived through stellar template synthesis modeling. The future evolution of this system is likely to be somewhere in between the two scenarios present above.  In addition, this amount of growth should be considered a lower limit, as massive nearby galaxies ($>10^{11}$\Msun), that lie beyond 150\,kpc, may also merge with the central galaxy. 

The crucial parameter that drives the evolution of such a system is the mass ratio of the central and satellite galaxies. If a galaxy is able to assemble quickly such that it is many times more massive than the general galaxy population at high redshift ($z>2$), a larger fraction of the stellar mass from its future mergers and close encounters will end up in an extended stellar halo. Thus whether or not a brightest cluster galaxy becomes a cD galaxy, may depend on its early-time formation history. The Spiderweb galaxy, with its large mass at $z=2.2$, is in a good position to build a substantial extended stellar halo by $z=0$.

\section{Concluding remarks}

We have used multi-wavelength high-resolution images to study the assembling radio galaxy, MRC1138-262, at $z\sim2.2$. We identified a galaxy population that lies at the same redshift and within a projected distance of 150\,kpc from the radio galaxy. Fitting stellar population models to the photometry of these galaxies we have examined the current distribution of stellar mass and ongoing star formation, and used analytic approximations to predict the future evolution of this galaxy. 

%\noindent 
1.~We identified 19 protocluster galaxies, comprising of a central massive galaxy ($10^{12}$\Msun) that is surrounded by 18 smaller satellite galaxies with masses between $10^{8}-10^{11}$\Msun. The central galaxy is already of a similar mass to local cD galaxies, but there is still room for growth as cD galaxies can have masses of $10^{13}$\Msun. The satellite galaxies  contribute between 10 and 50 per cent of the stellar mass of the whole system, and they dominate the star formation rate, containing $75\pm2$ per cent of the instantaneous star formation (dust-uncorrected).

%\noindent 
2.~The majority of the satellite galaxies have large star formation rates and are inferred to be gas-rich. The intrinsic star formation rates are uncertain because the red UV continuum slopes indicate dust obscuration. The high star formation rates of the satellite galaxies coupled with the presence of dust explains the spatially extended 850$\micron$ emission from MRC\,1138-262 \citep{Stevens2003}. The extended sub-mm emission most likely results from the sum of the compact star formation occurring in the radio and satellite galaxies rather than a system-wide starburst.  These results predict that a large fraction of the sub-mm flux should come from the west of the radio galaxy and high resolution sub-mm observations may be able to separate the emission into the various galaxy components.

%\noindent 
3.~We have compared the properties of the galaxies within 150\,kpc of MRC\,1138-262  to predictions of semi-analytic models. The models are able to reproduce the observed $K-$band galaxy luminosity function, and the derived stellar masses of the satellite galaxies. The  models predict  galaxy colours which are bluer  than the observations, and yet they do not match the observed high star formation rates. There may be a process, not accounted for in the models, which increases star formation when galaxies are situated in a dense merging environment, such as the Spiderweb system. 

%\noindent 
4.~Most of the satellite galaxies within 150\,kpc will accrete onto the central radio galaxy before $z=0$, increasing its mass by between $\sim$10 and 120 per cent. The merging timescales are long, and the vast majority of the galaxies will not merge for at least 300\,Myrs, by which time they will have completed many orbits. Gas is rapidly converted to stars as inferred from the large star formation rates. Therefore the final merging galaxies are likely to contain very little gas, and the mergers are likely to be dissipationless. 

%\noindent
5.~The tidal radii of the satellite galaxies are small, generally less than 10\,kpc, and in some cases are less than the half-light diameter of the galaxies. Therefore stars and gas are probably being stripped from the satellite galaxies. This stripping can occur at large distances from the central galaxy (more than 60\,kpc) and the stripped stars may form an extended stellar halo such as those that encompass local cD galaxies.

\section{Acknowledgments}
We would like to thank the anonymous referee for helping to improve and clarify the paper, Ian McCarthy for help with the Millennium simulations, Rychard Bouwens for help with the reduction of the NICMOS data, and Alex Saro for useful discussions. This research has been based on observations made with the NASA/ESA Hubble Space Telescope, obtained at the Space Telescope Science Institute, which is operated by the Association of Universities for Research in Astronomy, Inc., under NASA contract NAS 5-26555. These observations are associated with programs 10327 and 10404.
The Millennium Simulation databases used in this paper and the web
application providing online access to them were constructed as part of
the activities of the German Astrophysical Virtual Observatory.
NAH and GM acknowledge funding from the Royal Netherlands Academy of Arts and Sciences. JK is supported by the DFG, Sonderforschungsbereich (SFB) 439.
\bibliographystyle{mn2e}\bibliography{satellites,mn-jour}
\label{lastpage}
\clearpage
\end{document}